\documentclass[ALICE,manyauthors]{cernphprep}
\usepackage[english]{babel}
\usepackage{graphicx}
\usepackage{verbatim}
\usepackage{amsfonts}
\usepackage{makeidx}
\usepackage{color}
\usepackage{amsmath}
\usepackage{lineno}
\usepackage{epstopdf}
\usepackage{cite}
\usepackage{hyperref}

\newcommand{\pp}{{\rm pp}}

\newcommand{\sqrtsNN}{\sqrt{s_{\rm \scriptscriptstyle NN}}}

\newcommand{\av}[1]{\left\langle #1 \right\rangle}

\newcommand{\GeV}{\mathrm{GeV}}

\newcommand{\mev}{\mathrm{MeV}}
\newcommand{\gev}{\mathrm{GeV}}
\newcommand{\tev}{\mathrm{TeV}}

\newcommand{\cm}{\mathrm{cm}}

\newcommand{\mum}{\mathrm{\mu m}}

\newcommand{\ps}{\mathrm{ps}}

\newcommand{\PbPb}{\mbox{Pb--Pb}}

\newcommand{\RAA}{R_{\rm AA}}
\newcommand{\TAA}{T_{\rm AA}}

\newcommand{\vtwo}{v_{\rm 2}}

\newcommand{\pt}{\ensuremath{p_{\rm T}}}

\newcommand{\DtoKpi}{{\rm D^0\to K^-\pi^+}}
\newcommand{\DtoKpipi}{{\rm D^+\to K^-\pi^+\pi^+}}
\newcommand{\DstartoDpi}{{\rm D^{*+}\to D^0\pi^+}}

\newcommand{\Dzero}{{\rm D^0}}
\newcommand{\Dstar}{{\rm D^{*+}}}
\newcommand{\Dplus}{{\rm D^+}}

\newcommand{\dEdx}{{\rm d}E/{\rm d}x}

\renewcommand{\d}{\mathrm{d}}
\begin{document}

\begin{titlepage}

\PHnumber{2014-083}                 % required, obtained from PH
\PHdate{May 6, 2014}              % required

\title{Azimuthal anisotropy of $\rm D$ meson production\\ in Pb--Pb collisions at $\mathbf{\sqrtsNN=2.76~\tev}$}

\Collaboration{ALICE Collaboration%
         \thanks{See Appendix~\ref{app:collab} for the list of collaboration 
                      members}}
\ShortAuthor{ALICE Collaboration}
\ShortTitle{Azimuthal anisotropy of charm production in Pb--Pb collisions at $\sqrtsNN=2.76~\tev$}

\begin{abstract}
The production of the prompt charmed mesons $\Dzero$, $\Dplus$ and $\Dstar$
relative to the reaction plane  
was measured in Pb--Pb collisions at a 
centre-of-mass energy per nucleon--nucleon collision of $\sqrtsNN=2.76~\tev$ with the ALICE detector at the LHC. 
D mesons were reconstructed via their hadronic decays at central rapidity in the transverse momentum ($\pt$) interval 2--16~$\gev/c$.
The azimuthal anisotropy is quantified in terms of the second coefficient $v_2$ in a Fourier expansion 
of the D meson azimuthal distribution, and in terms of the nuclear modification factor $\RAA$, measured 
in the direction of the reaction plane and orthogonal to it.
The $v_2$ coefficient was measured with three different methods and in three centrality classes in the interval 0--50\%.
A positive $v_2$ is observed in mid-central collisions (30--50\% centrality class), with an mean value of $0.204_{-0.036}^{+0.099}$\,(tot.\,unc.) in the interval $2<\pt<6~\gev/c$, 
which decreases towards more central collisions (10--30\% and 0--10\% classes).
The positive $v_2$ is also reflected in the nuclear modification factor, which shows a 
stronger suppression in the direction orthogonal to the reaction plane for mid-central collisions.
The measurements are compared to theoretical calculations of charm quark transport and energy loss
in high-density strongly-interacting matter at high temperature.  
The models that include substantial elastic interactions with 
an expanding medium 
 provide a good description of the observed anisotropy. 
However, they are challenged 
to simultaneously describe
the strong suppression of high-$\pt$ yield of D mesons in central collisions and their azimuthal anisotropy in non-central collisions.
\end{abstract}

\end{titlepage}

\section{Introduction}
\label{sec:intro}

Collisions of heavy nuclei at ultra-relativistic energies are expected to lead to the formation of 
a high-density colour-deconfined state of strongly-interacting matter.
According to calculations of Quantum Chromo-Dynamics (QCD) on the lattice (see e.g.~\cite{karsch,Borsanyi,Bazavov,Petreczky}),
a phase transition to the Quark--Gluon Plasma (QGP) state can occur in these collisions, when conditions of 
 high energy density and temperature are reached. 
 Heavy quarks (charm and beauty), with large masses $m_{\rm c}\approx 1.3$ and  
$m_{\rm b}\approx 4.5~\gev/c^2$,
are produced in pairs predominantly at the initial stage of the collision~\cite{PBM} in hard scattering processes 
characterized by timescales shorter than the medium formation time.
They traverse the medium and interact with its constituents 
via both inelastic (medium-induced gluon radiation, i.e.\,radiative energy loss)~\cite{gyulassy,bdmps} and elastic (collisional)~\cite{thoma}
QCD processes.
Heavy-flavour hadrons
are thus effective probes of the properties of the medium formed in the collisions.

Compelling evidence for heavy-quark energy loss in strongly-interacting matter is provided by 
the observation of a modification of the transverse momentum ($\pt$) distributions of heavy-flavour hadrons.
This modification is quantified by the nuclear modification factor 
$R_{\rm AA}(\pt)=\d N_{\rm AA}/\d\pt \big/ \av{T_{\rm AA}} \d\sigma_{\rm pp}/\d\pt$,
where $\d N_{\rm AA}/\d\pt$ is the differential yield in nucleus--nucleus collisions in a given centrality class, 
 $\d\sigma_{\rm pp}/\d\pt$ is the cross section in pp collisions, and 
$\av{T_{\rm AA}}$ is the average nuclear overlap function~\cite{glauber}.
In central nucleus--nucleus collisions at RHIC and LHC energies,
$\RAA$ values significantly below unity were observed for heavy-flavour hadrons
with $\pt$ values larger than a few GeV/$c$~\cite{phenixRAAe,starRAAe,starDRAA,aliceDRAA,alicemuRAA,cmsJpsiRAA}.
A suppression by a factor up to 3--5 ($\RAA\approx 0.25$) at $\pt\simeq 5~\gev/c$ was measured 
in central collisions for 
inclusive electrons and muons from heavy-flavour hadron decays, both at RHIC ($\sqrtsNN=200~\gev$), by the PHENIX and STAR Collaborations~\cite{phenixRAAe,starRAAe},
and at the LHC ($\sqrtsNN=2.76~\tev$), by the ALICE Collaboration~\cite{alicemuRAA}.
At the LHC, the effect was also measured separately for charm, via D mesons by the ALICE Collaboration~\cite{aliceDRAA}, and for beauty,
 via non-prompt J/$\psi$ particles
from B hadron decays by the CMS Collaboration~\cite{cmsJpsiRAA}.

The D meson suppression at RHIC and at the LHC 
is described (see~\cite{starDRAA,aliceDRAA}) by model calculations that implement a combination of mechanisms of
heavy-quark interactions with the medium, via radiative and collisional processes, as well as in-medium formation and dissociation 
of charmed hadrons~\cite{vitevjet,whdg2011,beraudo,gossiaux,bamps,cujet,adsw}.
Model comparisons with more differential measurements can provide important insights
into the relevance of the various interaction mechanisms and the properties of the medium. In particular, the dependence of 
the partonic energy loss on the in-medium path length is expected to be different for each mechanism (linear for collisional processes~\cite{thoma} and close
to quadratic for radiative processes~\cite{bdmps}).
In addition, it is an open question whether low-momentum heavy quarks participate, through interactions with the medium, 
in the collective expansion of the system and whether they can reach thermal equilibrium with the medium 
constituents~\cite{batsouli,grecoetal}. It was also suggested that low-momentum heavy quarks could hadronize
not only via fragmentation in the vacuum, but also via the mechanism of recombination with other quarks from the medium~\cite{grecoetal,andronicetal}.

These questions can be addressed with
azimuthal anisotropy measurements of heavy-flavour hadron production with respect to the reaction plane,
defined by the beam axis and the impact parameter of the collision.
For non-central collisions, the two nuclei overlap in an approximately lenticular region, the short axis of which lies in the reaction plane.
Hard partons are produced at an early stage, when the geometrical anisotropy is not yet reduced by the system expansion. Therefore, partons emitted
in the direction of the reaction plane (in-plane) have, on average, a shorter in-medium path length than partons emitted orthogonally (out-of-plane),
leading {\it a priori} to a stronger high-$\pt$ suppression in the latter case.
In the low-momentum region, the in-medium interactions can also modify the parton emission directions, thus translating the initial spatial anisotropy into a momentum anisotropy of the final-state particles. 
Both effects cause a momentum anisotropy that can be characterized with the coefficients 
$v_n$ and the symmetry planes $\Psi_n$ of the Fourier expansion of the $\pt$-dependent particle distribution $\d^2N/\d\pt\d\varphi$ 
in azimuthal angle $\varphi$.
The elliptic flow is the second Fourier coefficient $v_2$, which can also be expressed as the average 
over all particles in all events of the angular correlation $\cos[2(\varphi-\Psi_2)]$.
The symmetry planes $\Psi_n$ for all harmonics would coincide with the
reaction plane if nuclei were spherically symmetric with a matter
density depending only on the distance from the centre
of the nucleus.
Due to fluctuations in the positions of the participant nucleons, the plane of 
symmetry fluctuates event-by-event around the reaction plane, independently for each harmonic, so that the $\Psi_n$ directions no longer coincide.

A path-length dependent energy loss, which gives a positive $v_2$, is considered to be
the dominant contribution to the azimuthal anisotropy of charged hadrons in the high $\pt$ region, above 8--10~GeV/$c$~\cite{highptv2a,highptv2b}.
At low $\pt$,
a large $v_2$ is considered as an evidence for the collective hydrodynamical expansion of the medium~\cite{Ollitrault,hydro}. 
Measurements of light-flavour hadron $v_2$ over a large $\pt$ range at RHIC and LHC are generally consistent with these expectations~\cite{PHENIXhighptv2,STARhighptv2,ALICEv2,ALICEhighptv2,ATLASv2,CMShighptv2,hydroVisc,whdg2011}.
In contrast to light quarks and gluons, which can be produced or annihilated 
during the entire evolution of the medium, heavy quarks are produced 
predominantly in initial hard scattering processes and their annihilation rate 
is small~\cite{PBM}.
Thus, the final state heavy-flavour hadrons at all transverse 
momenta originate from heavy quarks that experienced each stage of the 
system evolution.
High-momentum heavy quarks quenched by in-medium energy loss are shifted towards low momenta and, while 
participating in the collective expansion, they may
ultimately thermalize in the system.
In this context, the measurement of D meson $v_2$ is also important for the 
interpretation of recent results on J/$\psi$ anisotropy~\cite{jpsiv2}, because 
J/$\psi$ mesons formed from ${\rm c\overline{c}}$ recombination would inherit 
the azimuthal anisotropy of their constituent 
quarks~\cite{LiuXuZhuang,ZhaoEmerickRapp}. 

An azimuthal anisotropy in heavy-flavour production was observed in Au--Au collisions at RHIC with $v_2$ values of up to about $0.13$
for electrons from heavy-flavour decays~\cite{phenixHFEv2}. The measured asymmetry is reproduced by 
several models~\cite{gossiaux,bamps,beraudo,vanhees,moore,rappv2,rapp2014,lang,bass} implementing 
heavy-quark transport within a medium that 
undergoes a hydrodynamical expansion.
The transport properties, i.e.\ the diffusion coefficients, 
of heavy quarks in the medium can be related to its shear 
viscosity~\cite{moore}.
For LHC energies these models predict a large $v_2$ (in the range  0.10--0.20 in semi-central collisions) for $\rm D$ mesons at $\pt\approx 2$--$3~\gev/c$ and a decrease to a constant value $v_2\approx 0.05$ at high $\pt$.
The models described in Refs.~\cite{gossiaux,rappv2,rapp2014,lang,bass} include, at the hadronization 
stage, a contribution from the recombination of charm quarks 
with light quarks from the medium, which enhances $v_2$ at low $\pt$.

The measurement of the D meson $v_2$ in the centrality class 30--50\% 
in Pb--Pb collisions at $\sqrtsNN=2.76~\tev$, carried out using the ALICE detector,
 was presented in~\cite{Dv2Letter}.
The $v_2$ coefficient was found to be significantly larger than zero 
in the interval $2<\pt<6~\gev/c$ and comparable in magnitude with that of charged particles.

Here the measurement is extended to other centrality classes and accompanied with a study of the azimuthal dependence of the nuclear modification 
factor with respect to the reaction plane.  
The decays $\rm D^0\to K^-\pi^+$, $\rm D^+\to K^-\pi^+\pi^+$ and $\rm D^{*+}\to D^0\pi^+$ and charge
conjugates were reconstructed.
The  $v_2$ coefficient was measured with 
various methods in the centrality class 30--50\% as a function of $\pt$.
For the $\rm D^0$ meson, which has the largest statistical significance, the centrality dependence of $v_2$ in the range 0--50\% is presented 
and the anisotropy is also quantified in terms
of the nuclear modification factor $\RAA$ 
in two $90^\circ$-wide azimuthal intervals centred around the in-plane and out-of-plane directions. 

The experimental apparatus
is presented in Section~\ref{sec:detector}. The data analysis is described
in Section~\ref{sec:analysis}, including the 
data sample, the D meson reconstruction and the anisotropy measurement methods. 
Systematic uncertainties are discussed in Section~\ref{sec:syst}.
The results on $v_2$ and $\RAA$ are presented in Section~\ref{sec:results} and compared with 
model calculations in Section~\ref{sec:models}.

\section{Experimental apparatus}
\label{sec:detector}

The ALICE apparatus is described in~\cite{aliceJINST}. In this section, the characteristics of the detectors used for the D meson analyses are summarized. 
The $z$-axis of the ALICE coordinate system is defined by the beam direction, the $x$-axis lies in the horizontal plane and is pointing towards the centre 
of the LHC accelerator ring and the $y$-axis is pointing upward.

Charged-particle tracks are reconstructed in the central pseudo-rapidity\footnote{The pseudo-rapidity is defined as $\eta=-\ln(\tan\vartheta/2)$, where $\vartheta$ is the polar angle with respect to the $z$ axis.}
region ($|\eta|<0.9$) with the Time Projection Chamber (TPC) and the Inner Tracking System (ITS). For this analysis, charged hadron identification was performed using information from the TPC and the Time Of Flight (TOF) detectors.
These detectors are located inside a large solenoidal magnet that provides a field with a strength of $0.5~\mathrm{T}$, parallel to the beam direction.  
Two VZERO scintillator detectors, located in the forward and backward pseudo-rapidity regions, are used for online event triggering, 
collision centrality determination and, along with the Zero Degree Calorimeter (ZDC), for offline event selection.

The ITS~\cite{ITSalign} includes six cylindrical layers of silicon detectors surrounding the beam vacuum tube, at radial distances from the nominal beam line ranging from  3.9~cm for the innermost layer to 43~cm for the outermost one. 
The two innermost layers consist of Silicon Pixel Detectors (SPD) with a pixel size of $50\times425~\mum^2$ ($r\varphi\times z$, in cylindrical coordinates),
 providing an intrinsic spatial resolution of $12~\mum$ in $r\varphi$ and $100~\mum$ in $z$.  
The third and fourth layers use Silicon Drift Detectors (SDD) with an intrinsic spatial resolution of $35~\mum$ and $25~\mum$ in $r\varphi$ and $z$, respectively. 
The two outermost layers of the ITS contain double-sided Silicon Strip Detectors (SSD) with an intrinsic spatial resolution of $20~\mum$ in $r\varphi$ and $830~\mum$ in the $z$-direction. 
The alignment of the ITS sensor modules is crucial for the precise 
space point recontruction needed for the heavy-flavour analyses. It was performed
using survey information, cosmic-ray tracks and pp data. A detailed description of the employed methods can be found in~\cite{ITSalign}.
The effective spatial resolution along the most precise direction, $r\varphi$, is about 14, 40 and 25~$\mum$, for SPD, SDD and SSD, respectively~\cite{ITSalign,RossiVertex}. 

The TPC~\cite{TPCpaper} covers the pseudo-rapidity interval $|\eta|<0.9$ and extends in radius from $85~\cm$ to $247~\cm$. 
Charged-particle tracks are reconstructed and identified with up to 159 space points. 
The transverse momentum resolution for tracks reconstructed with the TPC and the ITS ranges from about 1\% at $\pt=1~\gev/c$ to about 2\% at $10~\gev/c$, both in pp and Pb--Pb collisions. The TPC also provides a measurement of the specific energy deposition $\dEdx$,
 with up to 159 samples. The truncated mean method, using only the lowest 60\% of the measured $\dEdx$ samples, gives a Gaussian distribution 
with a resolution (ratio of sigma over centroid) of about 6\%, which is slightly dependent on 
the track quality and on the detector occupancy. 

The TOF detector~\cite{newTOFpaper} is positioned at a radius of $370$--$399$~cm and it has the same pseudo-rapidity coverage as the TPC ($|\eta|<0.9$).  
The TOF provides an arrival time measurement for charged tracks with an overall resolution, including the measurement of the event start time, of about $80~\ps$ for pions and kaons at $\pt=1~\gev/c$ in the Pb--Pb collision centrality range used in this analysis~\cite{newTOFpaper}. 

The VZERO detector~\cite{vzero} consists of two arrays of scintillator counters covering the pseudo-rapidity regions $-3.7<\eta<-1.7$ (VZERO-C) and $2.8<\eta<5.1$ (VZERO-A).
Each array is composed of $8\times 4$ segments in the azimuthal and radial directions, respectively. 
This detector provides a low-bias interaction trigger (see Section~\ref{sec:sample}). 
For Pb--Pb collisions, the signal amplitude from its segments is used to classify events according to centrality, while the azimuthal segmentation allows for an estimation of the reaction plane.

The ZDCs are located on either side of the interaction point at $z\approx \pm 114$~m. 
The timing information from the neutron ZDCs was used to reject parasitic collisions between one of the two beams and residual nuclei present in the vacuum tube.

\section{Data analysis}
\label{sec:analysis}
\subsection{Data sample and event selection}
\label{sec:sample}

The analysis was performed on a data sample of $\PbPb$ collisions 
recorded in November and December 2011 
at a centre-of-mass energy per nucleon--nucleon collision of
$\sqrtsNN=2.76~\tev$.
The events were collected with an interaction trigger based on information 
from the VZERO detector, 
which required coincident signals recorded in the detectors at forward and backward pseudo-rapidities.
An online selection based on the VZERO signal amplitude was used
to enhance the sample of central and mid-central collisions through two 
separate trigger classes.
Events were further selected offline to remove background coming from parasitic
beam interactions by using the time information provided by the 
VZERO and the neutron ZDC detectors. 
Only events with a reconstructed interaction point
(primary vertex), determined by extrapolating charged-particle tracks, within $\pm 10$~cm from the centre
of the detector along the beam line were used in the analysis.

Collisions were classified in centrality classes, 
determined from the sum of the amplitudes of the signals in the VZERO detector
and defined in terms of percentiles of the total hadronic Pb-Pb cross section.
In order to relate the centrality classes to the collision geometry, 
the distribution of the VZERO summed amplitudes
was fitted by a model based on the Glauber approach for the geometrical 
description of the nuclear collision~\cite{glauber} complemented
by a two-component model for particle production~\cite{centrality}.
The centrality classes used in the analysis are 
reported in Table~\ref{tab:stats}, together with the number of events in each class and the corresponding
integrated luminosity.

\begin{table}[!t]
\centering
\caption{Number of events and integrated luminosity for the considered centrality classes, expressed as percentiles of the hadronic cross section.
The uncertainty on the integrated luminosity derives from the uncertainty of the hadronic Pb--Pb cross section from the Glauber model~\cite{glauber,centrality}.}
\begin{tabular}{|ccc|} 
\hline 
\rule[-0.3cm]{0cm}{0.8cm} Centrality class &  $N_{\rm events}$ & $L_{\rm int}~(\mu{\rm b}^{-1})$\\
\hline
\phantom{0}0--10\%  & $16.0\times 10^6$  &  $20.9\pm 0.7$ \\
10--30\%  & \phantom{1}$9.5\times 10^6$  & \phantom{2}$6.2\pm 0.2$  \\
30--50\%  & \phantom{1}$9.5\times 10^6$  & \phantom{2}$6.2\pm 0.2$ \\
\hline
\end{tabular}
\label{tab:stats}
\end{table}

\subsection{D meson reconstruction}
\label{sec:reco}

The $\Dzero$, $\Dplus$ and $\Dstar$ mesons and their antiparticles were reconstructed in the rapidity interval $|y|<0.8$ via their hadronic decay channels $\DtoKpi$ (with branching ratio, BR, of $3.88 \pm 0.05\%$), \mbox{$\DtoKpipi$} (BR = $9.13 \pm 0.19\%$), and $\DstartoDpi$ (BR = $67.7\pm 0.5\%$) and their corresponding charge conjugates~\cite{PDG}.
The $\Dzero$ and $\Dplus$ mesons decay weakly with mean proper decay lengths ($c\tau$) of approximately 123 and 312~$\mum$~\cite{PDG}. 
The $\Dstar$ meson decays strongly at the primary vertex.

$\Dzero$ and $\Dplus$ candidates were defined from pairs and triplets of tracks within the fiducial acceptance $|\eta|<0.8$, selected by requiring at least 70 associated space points in the TPC, $\chi^2/\rm ndf <2$ for the momentum fit, and at least two associated hits in the ITS, with at least one of them in the SPD. A transverse momentum threshold $ \pt > 0.4~\GeV/c$ was applied in order to reduce the combinatorial background.
$\Dstar$ candidates were obtained by combining the $\Dzero$ candidates with 
tracks selected with the same requirements as described above, but with a lower transverse momentum threshold $\pt > 0.1~\GeV/c$ and at least three associated 
hits in the ITS, with at least one of them in the SPD. 
The lower $\pt$ threshold was used because the momentum of the 
pions from $\Dstar$ decays is typically low, as a consequence of the small mass 
difference between $\Dstar$ and $\Dzero$.

The selection of tracks with $|\eta|<0.8$ introduces a steep drop in the acceptance of D mesons for rapidities larger than 0.7--0.8, depending on $\pt$.  
A fiducial acceptance region was, therefore, defined as: $|y| < y_{\rm fid}(\pt)$, with $y_{\rm fid}(\pt)$ increasing from 0.7 to 0.8 in $2 < \pt < 5~ \GeV/c$ and taking a constant value of 0.8 for $\pt > 5 ~\GeV/c$~\cite{aliceDRAA}. The D meson $\vtwo$ results are not expected to be affected by this small variation in rapidity acceptance.

The D meson yields were measured with an invariant mass analysis of reconstructed decays, using  
kinematic and geometrical selection criteria, and particle identification (PID).
The selection of $\Dzero$ and $\Dplus$ decays was based on the reconstruction of secondary vertices with a separation of a few hundred microns from primary vertex. In the case of the $\Dstar$ decay, the secondary vertex of the produced  $\Dzero$ was reconstructed. 
The coordinates of the primary vertex and of the secondary vertices, as well as the corresponding covariance matrices,
were computed using a $\chi^2$ minimization method~\cite{Dpp7}.

The selection strategy is the same as in previous pp~\cite{Dpp7,Dpp276} and Pb--Pb~\cite{aliceDRAA} analyses. It exploits the 
displacement of the decay tracks from the primary vertex (transverse impact parameter, $d_0$), the
separation between the secondary and primary vertices (decay length, $L$) and the pointing of the reconstructed meson momentum to the primary vertex.

The transverse impact parameter $d_0$ of a given track is defined as the signed distance of closest approach of the extrapolated track to the primary vertex in the $(x,y)$ plane. The sign of $d_0$ is attributed based on the position of the primary vertex with respect to the curve of the $(x,y)$ projection of the track. 
In Pb--Pb collisions, the impact parameter resolution in the transverse direction  is better than $65~\mum$ for tracks with a transverse momentum larger than $1~\gev/c$ and reaches $20~\mum$ 
for $\pt>20~\gev/c$~\cite{aliceDRAA}. This includes the contribution from the primary vertex precision, which is better than 
$10~\mum$ in the central and semi-central Pb--Pb events used in this analysis.
The impact parameter measurement is significantly less precise along the longitudinal direction, e.g.\,$170~\mum$ at $\pt=1~\gev/c$. 

A pointing condition was applied via a selection on the angle $\vartheta_{\rm pointing}$ between the direction of the reconstructed momentum of the candidate and the straight line connecting the primary and secondary vertices.
For $\PbPb$ collisions, two additional selection variables were introduced with respect to $\pp$ analyses, namely the projection of the pointing angle and of the decay length onto the transverse plane ($\vartheta^{xy}_{\rm pointing}$ and $L^{xy}$). 
The selection requirements were tuned so as to provide a large statistical significance for the signal and to keep the selection efficiency as high as possible. 
The chosen selection values depend on the $\pt$ of the D meson and become more stringent from peripheral to central collisions.

The selection criteria for the centrality class 30--50\% are described in the following. 
The $\Dzero$ candidates were selected by requiring the decay tracks to have an impact parameter significance $|d_0|/\sigma_{ d_0} > 0.5$
($\sigma_{ d_0}$ is the uncertainty on the track impact parameter),
and to form a secondary vertex with a track-to-track distance of closest 
approach smaller than $250$--$300~\mum$, depending on $\pt$, and a decay length larger than  $100~\mum$. 
The product of the decay track impact parameters, 
which are of opposite sign for well-displaced signal topologies,
was required to be below $-(200~\mum)^2$ at low $\pt$ (2--3~$\gev/c$) and below $-(120~\mum)^2$ for high $\pt$ candidates (12--16~$\gev/c$), with a smooth variation between these values in 2--12~$\gev/c$.  A significance of the projection of the decay length in the transverse plane  $L^{xy}/\sigma_{ L^{xy}}$ 
(where $\sigma_{ L^{xy}}$ is the uncertainty on $L^{xy}$)
larger than 5 was also required. A selection on the angle $\vartheta^*$ between the kaon momentum in the $\Dzero$ rest frame and the boost direction was used to reduce the contamination from background candidates that do not represent real two-body decays and typically have large values of $ | \cos \vartheta^*|$. The selection $ | \cos \vartheta^*|<0.8$ was applied. The pointing of the $\Dzero$ momentum to the primary vertex was implemented by requiring $\cos \vartheta_{\rm pointing} > 0.95$ and $\cos \vartheta^{xy}_{\rm pointing} > 0.998$ at low $\pt$ (2--3~$\gev/c$). Since the background is lower at high $\pt$, the cuts were progressively made less stringent for increasing $\pt$. 
In the 0--10\% and 10--30\% centrality classes the combinatorial background is larger than in 30--50\%. Therefore, the selections were made more
stringent and they are similar to those used for the 0--20\% centrality class in~\cite{aliceDRAA}.

The $\Dplus$ candidates were selected by requiring a decay length larger 
than $1200$--$1600~\mum$, depending on $\pt$, and $\cos\vartheta_{\rm pointing}$ larger 
than 0.998 (0.990) in the $\pt$ interval 3--4 (8--12)~$\gev/c$, with a smooth variation in-between.
Further requirements to reduce the combinatorial background were 
$\cos\vartheta_{\rm pointing}^{xy}> 0.993$--0.998 and $L^{xy}/\sigma_{L^{xy}}>9$--11, 
depending on the candidate $\pt$. 
In general, the $\Dplus$ selection criteria are more stringent than those of the $\Dzero$ 
because of the larger combinatorial background.

In the $\Dstar$ analysis, the selection of the decay $\Dzero$ candidates was similar
to that used for the $\Dzero$ analysis. Only $\Dzero$ candidates with invariant mass within $2.5\,\sigma$ of the PDG mass value~\cite{PDG} were used, where $\sigma$ is the $\pt$-dependent Gaussian sigma of the invariant mass distribution observed in data.
The decay pion was selected with the same track quality criteria as for the $\Dzero$ and $\Dplus$ decay tracks.

Pions and kaons 
were identified with the TPC and TOF detectors,
on the basis of the difference, expressed in units of 
the resolution ($\sigma$), between the measured signal and that expected for 
the considered particle species.
Compatibility regions at $\pm3\,\sigma$ around the expected mean energy deposition 
$\dEdx$ and time-of-flight were used.
Tracks without a TOF signal were identified using only the TPC information.
This particle identification strategy provided a reduction by a factor of about three of the combinatorial background in the low-$\pt$ range, while preserving most of the signal (see Section~\ref{sec:raavsep}).

The $\Dzero$ and $\Dplus$ raw yields were obtained with a fit to the invariant mass $M$ distribution of the D meson candidates.
For the $\Dstar$ signal the mass difference $\Delta M=M({\rm K}^-\pi^+\pi^+) - M({\rm K}^-\pi^+)$ was considered. 
The fit function is the sum of a Gaussian to describe the signal and a term describing the background, which is an exponential for $\Dzero$ and $\Dplus$ 
and has the form $f(\Delta M)=a\,(\Delta M-m_\pi)^b$ for the $\Dstar$, where $m_\pi$ is the charged pion mass and $a$ and $b$ are free parameters. 
The centroids and the widths of the Gaussian functions were found to be 
in agreement, respectively, with the D meson PDG mass values~\cite{PDG} and with the simulation results, 
confirming that the background fluctuations were not causing a distortion in the signal line shape. 
An example of invariant mass distributions will be shown in 
Section~\ref{sec:methods}.

\subsection{Azimuthal anisotropy analysis methods}
\label{sec:methods}

The $\pt$-differential azimuthal distribution of produced particles can be 
described by a Fourier series:
\begin{equation} \label{eq:fourier}
\frac{\d^2 N}{\d\varphi\d\pt}=\frac{\d N}{2\pi\d\pt}\left[1+2\sum\limits^{\infty}_{n=1}v_n(\pt)\,\cos n(\varphi-\Psi_n)\right]\,,
\end{equation}
where $\Psi_n$ is the initial state spatial plane of symmetry of the $n$-th harmonic, 
defined by the geometrical distribution of the 
nucleons participating in the collision.
In order to determine the second harmonic coefficient $\vtwo$, the $\vec{Q}$ vector 
\begin{equation}\label{eq:qvector}
 \vec{Q}={\sum_{i=1}^{N} w_i \cos 2\varphi_i \choose \sum_{i=1}^{N} w_i \sin 2\varphi_i} \qquad
\end{equation}
is defined from the azimuthal distribution of charged particles, where
$\varphi_i$ are the azimuthal angles and $N$ is the multiplicity of
charged particles. The weights $w_i$ are discussed later in the
text. The charged particles used for the $\vec{Q}$ vector determination are
indicated in the following as reference particles (RFP).
The azimuthal angle of the $\vec{Q}$ vector
\begin{equation}
\psi_2 = \dfrac{1}{2} \tan^{-1} \left(\dfrac{Q_{y}}{Q_{x}}\right)
\end{equation}
is called event plane angle and it is an estimate of the second harmonic symmetry plane $\Psi_2$.

The event plane (EP)~\cite{ref:ep}, scalar product (SP)~\cite{SP} and two-particle cumulant methods~\cite{ref:BSVCumulants} were used to measure the D meson elliptic flow.

\begin{figure*}[!t]
\begin{center}
\includegraphics[width=0.49\textwidth]{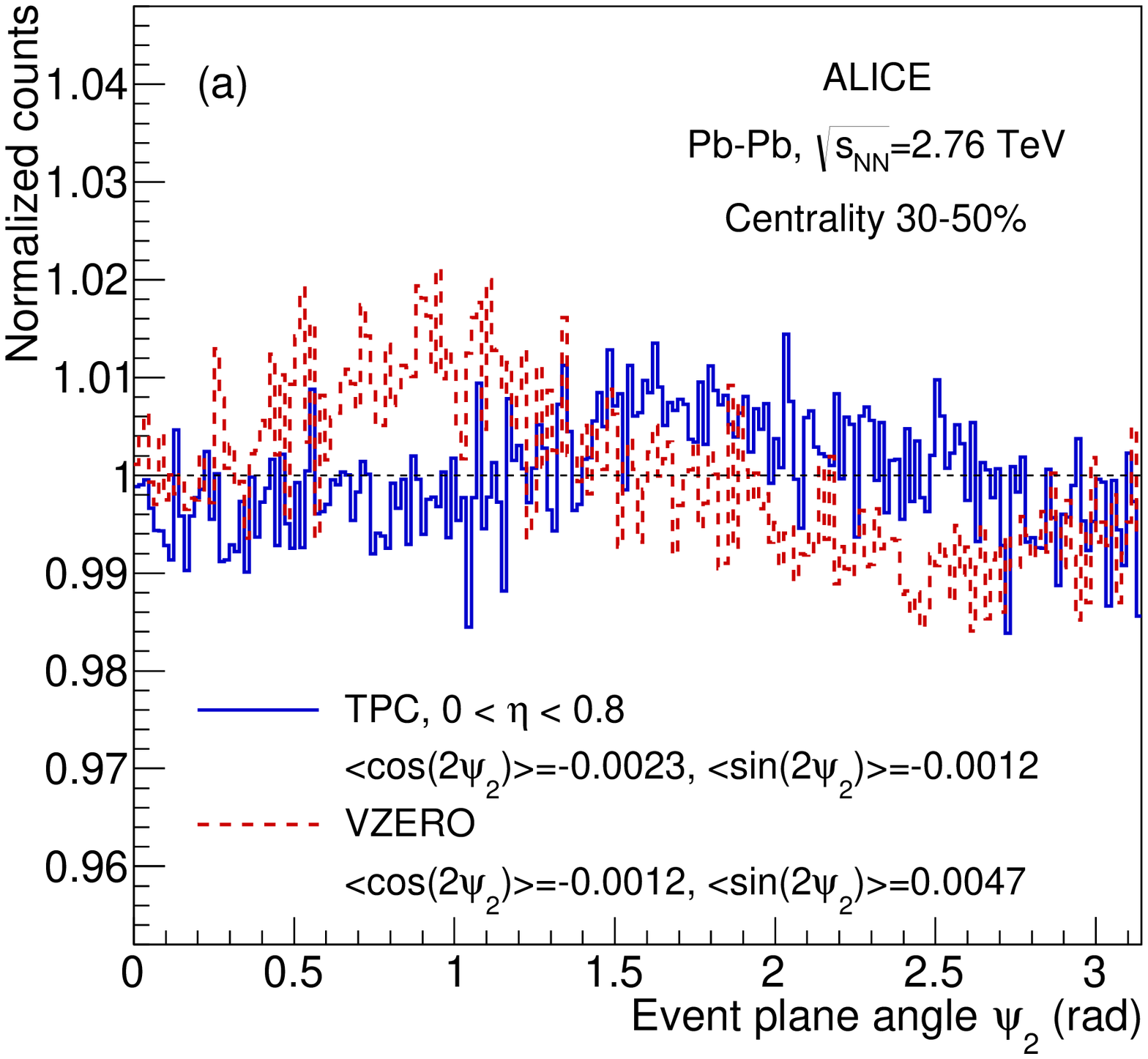}
\includegraphics[width=0.49\textwidth]{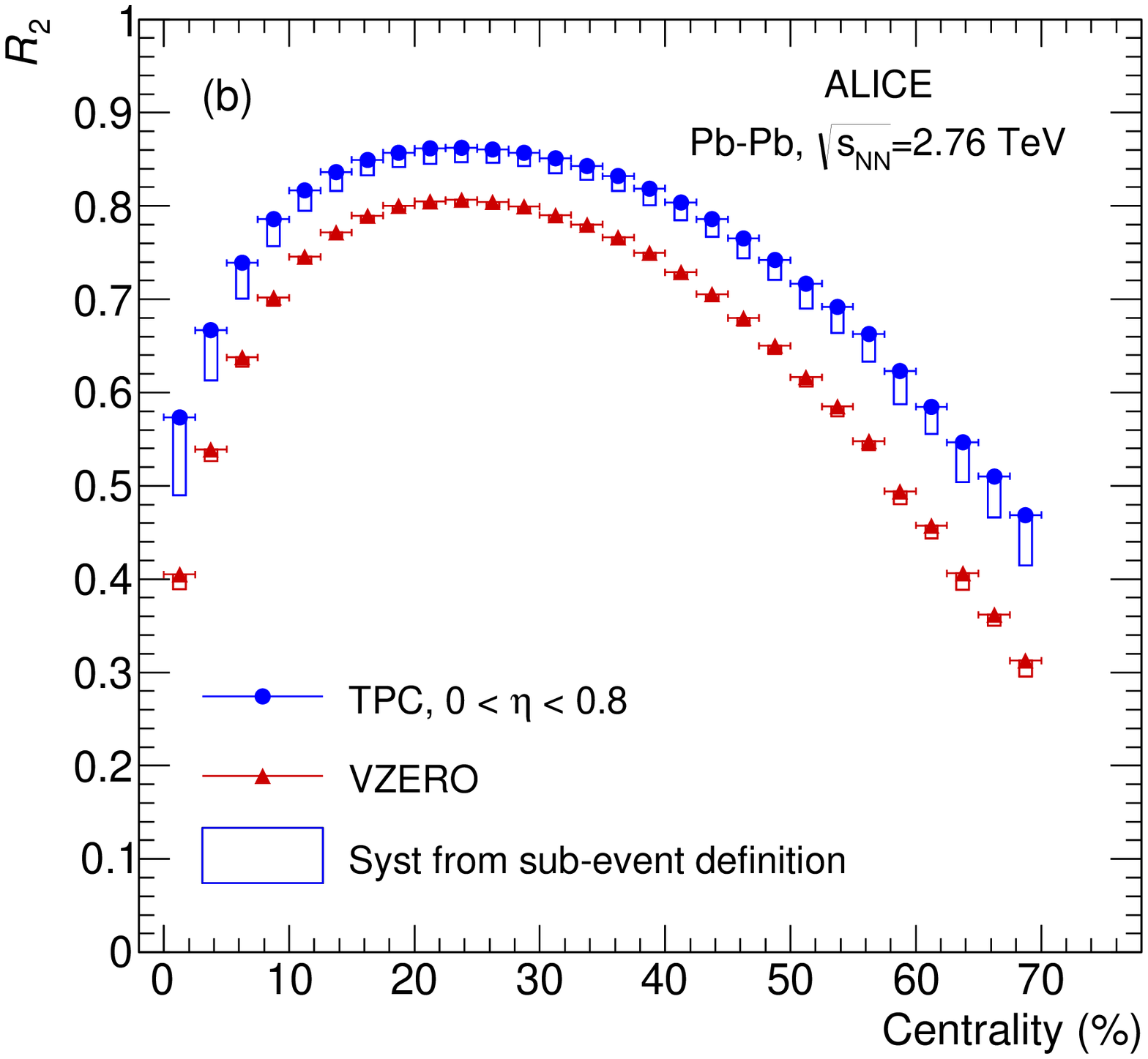}
\caption{(Color online) (a) Distribution of event plane angle $\psi_2$, estimated from TPC tracks with $0<\eta<0.8$ (solid line) or with the VZERO detector signals (dashed line) in the centrality range 30--50\%. The distributions are normalized by their integral. (b) Event plane resolution correction factor $R_2$ as a function of centrality for the TPC and VZERO detectors. The boxes represent the systematic uncertainties estimated from the variation of $R_2$ when changing the sub-events used for its determination. 
}
\label{fig:planes}
\end{center}
\end{figure*}

The charged particle tracks used for the $\vec{Q}$ vector determination were selected with the following criteria: 
at least 50 associated space points in the TPC;
$\chi^2/{\rm ndf}<2$ for the momentum fit in the TPC; 
a distance of closest approach to the primary vertex smaller than 3.2~cm in $z$ and 2.4~cm in the $(x, y)$ plane. 
In order to minimize the non-uniformities in the 
azimuthal acceptance, no requirement was applied on the number of ITS points 
associated to the track.
To avoid auto-correlations between the D meson candidates and the event plane angles, 
the $\vec{Q}$ vector was calculated for each candidate excluding 
from the set of reference particles the tracks used to form that 
particular candidate.
Tracks with $\pt>150~\mev/c$ were considered and  
the pseudo-rapidity interval was limited to the positive region $0<\eta<0.8$, where the TPC acceptance and 
efficiency were more uniform as function of the azimuthal angle for this data set.
The remaining azimuthal non-uniformity was corrected for using weights $w_i$ in Eq.~(\ref{eq:qvector}), defined as the inverse of
the $\varphi$ distribution of charged particles used for the $\vec{Q}$ vector determination, $1/({\rm d}N/{\rm d}\varphi_i)$, multiplied by a function $f(\pt)=\big\{^{\pt/\gev/c,~\pt<2~\gev/c}_{2,~\pt\ge2~\gev/c}$. This function mimics the $\pt$-dependence of the charged particle $v_2$ and it improves the estimate of $\Psi_2$ by 
enhancing the contribution of particles with a stronger flow signal (see e.g.\,Ref.~\cite{ALICEhighptv2}).
The distribution of the event plane angle $\psi_2$ obtained for this set of reference particles is shown in Fig.~\ref{fig:planes}~(a), for the centrality range 30--50\%. The distribution, divided by its integral, exhibits a residual non-uniformity below 1\%.

An additional study was performed with the $\vec{Q}$ vector determined from the azimuthal distribution of signals in the segments of the VZERO 
detectors, which are sensitive to particles produced at forward and backward 
rapidities.
The $\vec{Q}$ vector was calculated with Eq.~(\ref{eq:qvector}), with the sum
running over the eight azimuthal sectors of each VZERO detector,  
where $\varphi_i$ was defined by the central azimuth of the $i$-th sector, and 
$w_i$ equal to the signal amplitude in the $i$-th sector for the selected event, which is 
proportional to the number of charged particles crossing the sector.
Non-uniformities in the VZERO acceptance and efficiency were corrected for
using the procedure described in~\cite{nonunifcorr}.
The residual non-uniformity is about 1\%, as shown in Fig.~\ref{fig:planes}~(a).

For the event plane method,
the measured anisotropy $v_2^{\rm obs}$ was divided by the event plane resolution correction factor $R_2$ according to the equation $v_2=v_2^{\rm obs}/ R_2$, with 
$R_2$ being smaller than one.
This resolution depends on the multiplicity and $v_2$ of the RFP~\cite{ref:ep}. 
For the event plane computed using TPC tracks,  
$R_2$ was determined from the correlation of the
event plane angles reconstructed from RFP in the two sides of the TPC, 
$-0.8<\eta<0$ and $0<\eta<0.8$, i.e. two samples of tracks (called sub-events) with similar multiplicity and $v_2$.
$R_2$ is shown in Fig.~\ref{fig:planes}~(b) as a function of collision centrality.
The average $R_2$ values in the three centrality classes used in this analysis 
are 
$0.6953$ (0--10\%), 
$0.8503$ (10--30\%) and 
$0.8059$ (30--50\%).   The statistical uncertainty on $R_2$ is negligible ($\sim 10^{-4}$).
The systematic uncertainty on $R_2$ was estimated by using the three-sub-event method described in~\cite{ref:threesubeventmethod}. In this case, 
the event planes reconstructed in the TPC ($0<\eta<0.8$), VZERO-A ($2.8<\eta<5.1$) and VZERO-C ($-3.7<\eta<-1.7$) were used. 
This method yielded $R_2$ values smaller than those obtained from the two-sub-events method 
by 6.9\%, 2.0\% and 2.3\% for the centrality classes 0--10\%, 10--30\% and 30--50\%. 
A part of this difference can be 
attributed to the presence of short-range non-flow 
correlations that are suppressed when the three sub-events with a pseudo-rapidity 
gap are used. Non-flow correlations can originate  from resonance or cascade-like decays and from jets.
The resolution of the event plane determined from the VZERO detector (summing 
the signals in VZERO-A and VZERO-C) is also shown in Fig.~\ref{fig:planes}~(b).
In this case, $R_2$ was measured with three sub-events, namely the signals in
the VZERO detector (both A and C sides) and the tracks in the positive and negative $\eta$ regions of the TPC.
The systematic uncertainty was estimated from the difference with
the results obtained with two TPC sub-events separated by 0.4 units in pseudo-rapidity ($\eta$ gap).
The event plane determination has a poorer resolution with the VZERO detector
than with the TPC tracks. As a consequence, the $v_2$ measurement is expected to be more precise
with the TPC event plane.

\begin{figure*}[!t]
\begin{center}
\includegraphics[width=\textwidth]{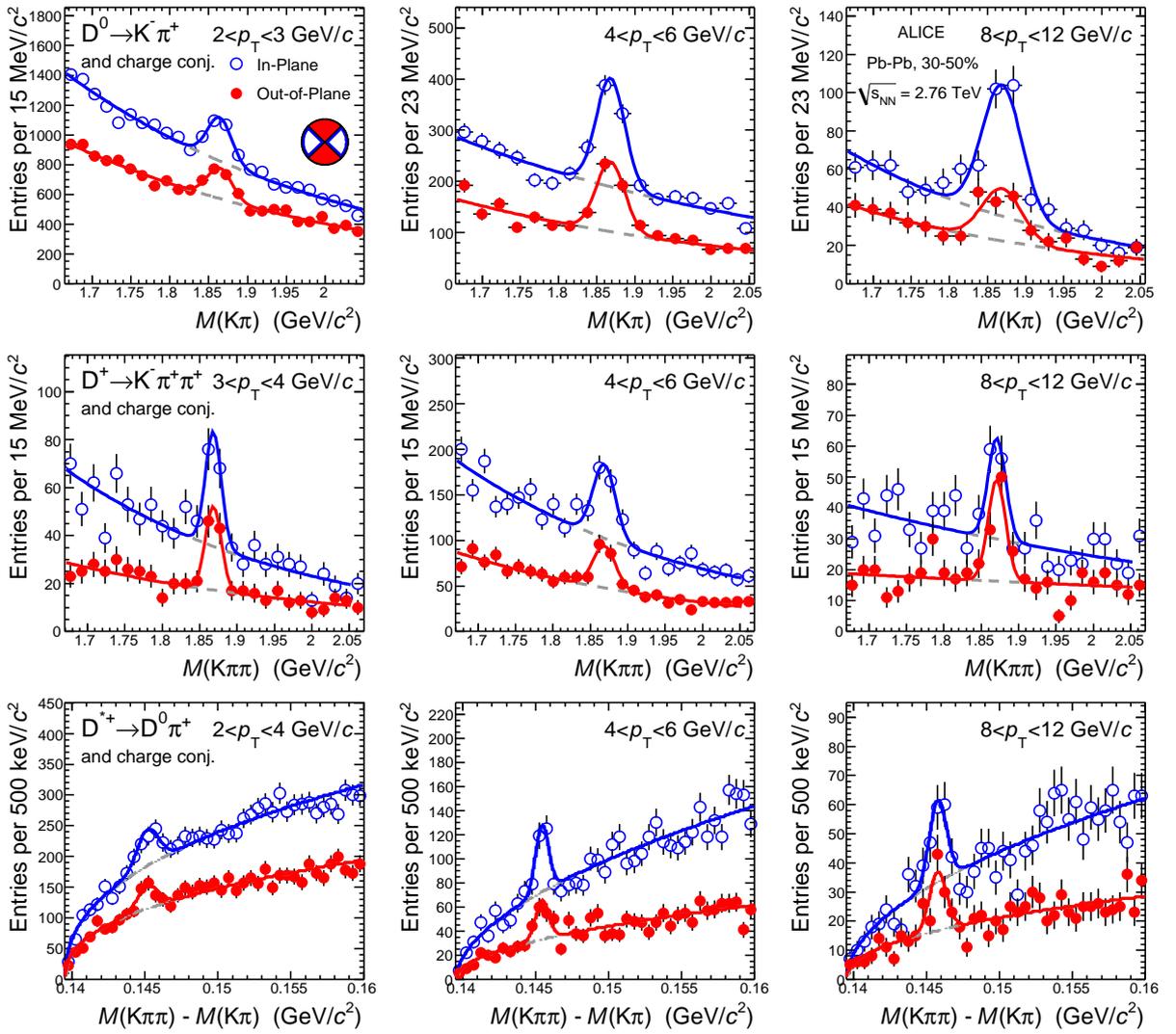}
\caption{(Color online) Distributions of the invariant mass for $\Dzero$ (upper panels) and $\Dplus$ (central panels) candidates and of the mass difference for $\Dstar$ candidates (lower panels) in the two $\Delta\varphi$ intervals used in the event plane method, for Pb--Pb collisions in the 30--50\% centrality class. The rapidity interval is $|y|<y_{\rm fid}$ (see Section~\ref{sec:reco} for details). For each meson species three $\pt$ intervals are shown, along with the fits used to extract the signal yield. The definition of the two $\Delta\varphi$ intervals is sketched in the top-left panel.}
\label{fig:MassPlots3050EP}
\end{center}
\end{figure*}

In the event plane method, the D meson yield was measured in two $90^\circ$-wide intervals of $\Delta\varphi=\varphi_{\rm D}-\psi_2$: {\it in-plane} 
($-\frac{\pi}{4}<\Delta\varphi\le\frac{\pi}{4}$ and $\frac{3\pi}{4}<\Delta\varphi\le\frac{5\pi}{4}$) and {\it out-of-plane} 
($\frac{\pi}{4}<\Delta\varphi\le\frac{3\pi}{4}$ and $\frac{5\pi}{4}<\Delta\varphi\le\frac{7\pi}{4}$). 
$\varphi_{\rm D}$ is defined as the azimuthal angle of the D meson momentum vector at the primary vertex.
The invariant mass distributions for the three meson species  are shown in Fig.~\ref{fig:MassPlots3050EP} in three $\pt$ intervals for the 30--50\% centrality class, along with the fits used for the yield estimation
(Section~\ref{sec:reco}). 
When  fitting the invariant mass distribution in the two $\Delta\varphi$
intervals, the centroid and the width of the Gaussian functions were fixed, for each meson species and 
for each $\pt$ interval, to those obtained from a fit to the invariant mass 
distribution integrated over $\varphi$, where the statistical 
significance of the signal is larger.

Integrating Eq.~(\ref{eq:fourier}) and including the correction for the event plane resolution $1/R_2$ yields:
\begin{equation} 
\label{eq:twobins} 
v_2\{{\rm EP}\} =\frac{1}{R_2}\frac{\pi}{4}\frac{N_{\textnormal{in-plane}}-N_{\textnormal{out-of-plane}}}{N_{\textnormal{in-plane}}+N_{\textnormal{out-of-plane}}}\,.
\end{equation}

The contribution of higher harmonics to the 
$v_2$ value calculated with this equation can be evaluated by integrating the 
corresponding terms of the Fourier series.
All odd harmonics, as well as $v_4$ and $v_8$, induce the same 
average contribution to $N_{\rm in\mbox{-}plane}$ and 
$N_{\rm out\mbox{-}of\mbox{-}plane}$ 
due to symmetry, and therefore they do not affect $v_2$ calculated with Eq.~(\ref{eq:twobins}).
The contribution of $v_6$, $v_{10}$ and higher harmonics is assumed to be
negligible based on the values measured for light-flavour hadrons~\cite{ALICEcorrel,ATLASvn}. 

\begin{figure*}[!t]
\begin{center}
\includegraphics[width=0.49\textwidth]{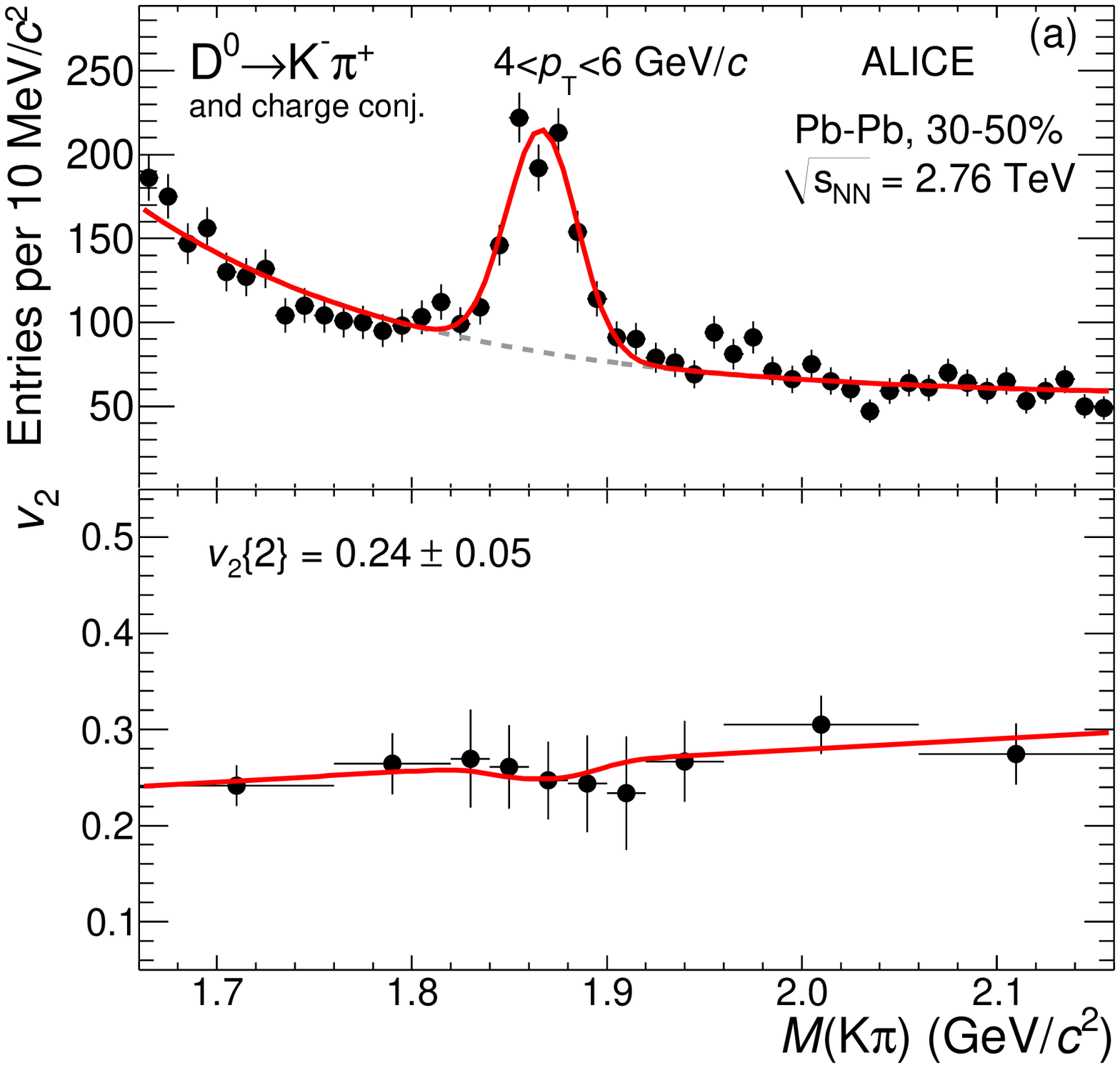}
\includegraphics[width=0.49\textwidth]{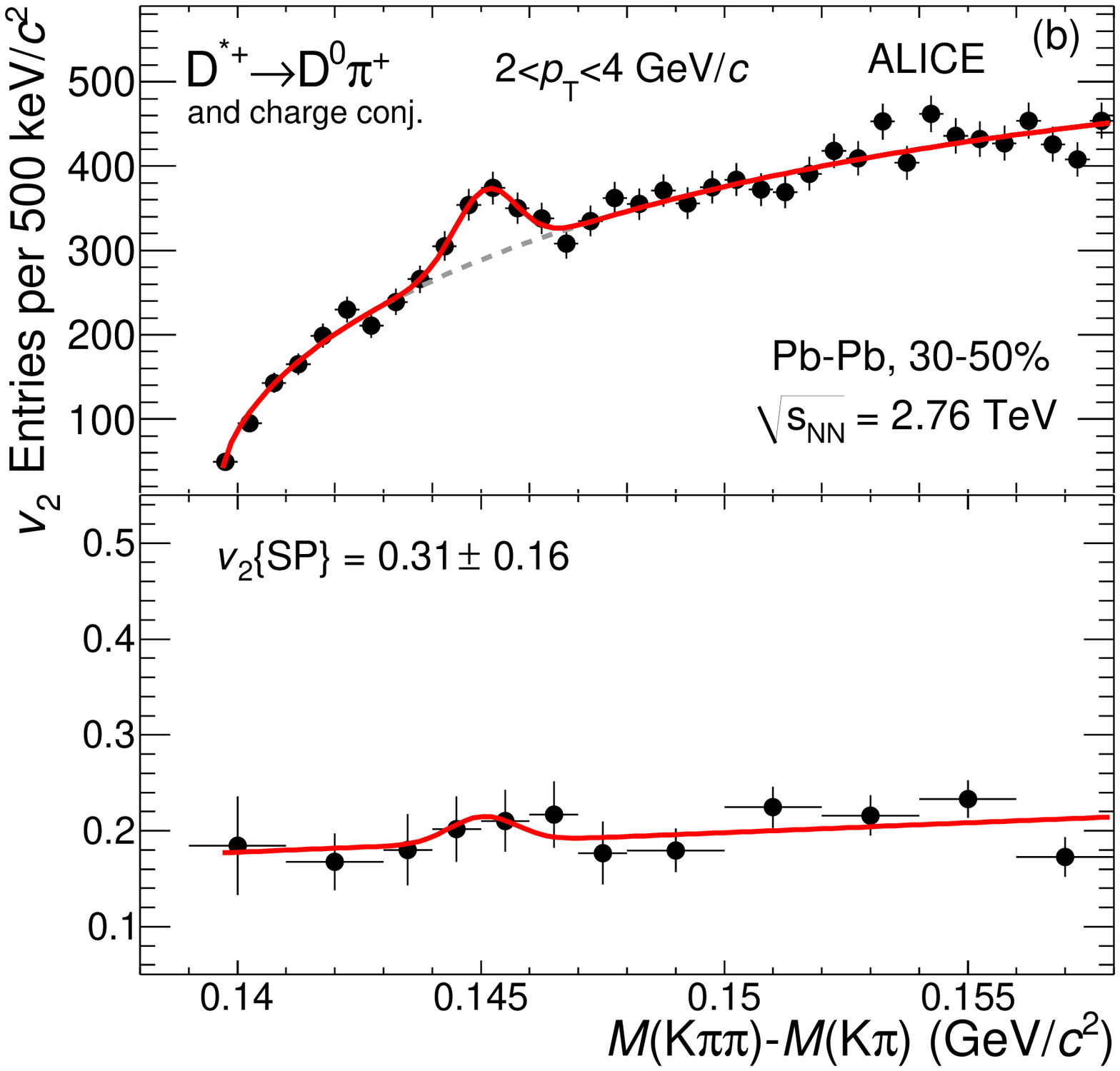}
\caption{(Color online) Examples of $v_2$ extraction with two-particle correlation methods in a selected $\pt$ interval for Pb--Pb collisions in the 30--50\% centrality range: the two-particle cumulants method for $\rm D^0$ (a) and the scalar product  method for $\rm D^{*+}$ (b). The lower panels report the D meson $v_2$ values obtained with the simultaneous fit procedure, as described in the text. The rapidity interval is $|y|<y_{\rm fid}$ (see Section~\ref{sec:reco} for details).}
\label{fig:QCSPfit}
\end{center}
\end{figure*}

The measurement of the elliptic flow with the scalar product method is given by~\cite{ref:ep}:
\begin{equation}
\label{eq:sp}
v_2\{{\rm SP}\} = 
\frac{1}{2}\left(
\frac
{\big\langle \vec{u}_a \cdot \frac{\vec{Q}_b}{N_b}\big\rangle}
{\sqrt{\big\langle \frac{\vec{Q}_a}{N_a} \cdot \frac{\vec{Q}_b}{N_b} \big\rangle}}
+
\frac
{\big\langle  \vec{u}_b \cdot \frac{\vec{Q}_a}{N_a}\big\rangle}
{\sqrt{\big\langle \frac{\vec{Q}_a}{N_a} \cdot \frac{\vec{Q}_b}{N_b} \big\rangle}}
\right)\,,
\end{equation}
where $\langle~\rangle$ indicates an average over D meson candidates 
in all events.
The vector $\vec{u}$ is defined as $\vec{u}=(\cos 2\varphi_{\rm D},\sin 2\varphi_{\rm D})$, where $\varphi_{\rm D}$ the D meson candidate momentum azimuthal direction.
The $\vec{Q}_{a,b}$ and $\vec{u}_{a,b}$ vectors were computed from charged particles and D meson candidates, respectively,
in two separate pseudo-rapidity regions: 
$a$) $0< \eta < 0.8$ and $b$) $-0.8< \eta < 0$.
The elliptic flow was computed by correlating D mesons from the positive 
$\eta$ region with the charged particles in the negative $\eta$ region, and vice versa.
This separation in pseudo-rapidity
suppresses two-particle correlations at short distance that are due to decays (${\rm D^* \to D} +X$ and ${\rm B \to D^{(*)}} +X$).  
The denominator in Eq.~(\ref{eq:sp}) plays a similar role as the resolution correction in the event plane method.
Since the resolution is proportional to the number of used RFP, the vectors $\vec{Q}_a$ and $\vec{Q}_b$  were normalized by $N_a$ and $N_b$, respectively, before averaging over all events. 
The azimuthal non-uniformity of the TPC response, which results in non-zero
average values of $\vec{Q}_a$ and $\vec{Q}_b$, was corrected for using a re-centering procedure~\cite{ref:ep}:
$\vec{Q}_{a,b}'=\vec{Q}_{a,b}-\langle \vec{Q}_{a,b} \rangle$.

The two-particle cumulant is defined by the equation~\cite{ref:BSVCumulants,ref:AliCumulants1,ref:AliCumulants2}:
\begin{equation}
v_2\{2\}=\frac{ \big\langle \vec{u}\cdot \frac{\vec{Q}}{N} \big\rangle }{\sqrt{ \big\langle \frac{\vec{Q}_a}{N_a} \cdot \frac{\vec{Q}_b}{N_b} } \big\rangle }  \,.
\end{equation}
For this method,  the  
 azimuthal non-uniformity of the detector acceptance and efficiency was corrected for with the aforementioned re-centering procedure.
In contrast to the scalar product method, there is no pseudo-rapidity gap between the
D mesons and the RFP for the two-particle cumulant method.

For both the scalar product and two-particle cumulant methods, the $v_2$ of D meson candidates was computed 
in narrow intervals of invariant mass $M$ for $\Dzero$ and $\Dplus$ and mass difference $\Delta M$ for the $\rm D^{*+}$. In each invariant mass interval, 
the measured $v_2$ is the weighted average of the D meson $v_2$ ($v_2^{\rm S}$) and the background $v_2$ ($v_2^{\rm B}$) with the weights given by the relative fractions of signal (S) and background (B) in that interval. In order to 
 extract the values of  $v_2^{\rm S}$ and $v_2^{\rm B}$, a simultaneous fit of the distributions of counts and $v_2$ as a function of invariant mass $M$ was performed. 
The invariant mass distribution was fitted with a sum of two terms for signal and background, as explained in Section~\ref{sec:reco}. The $v_2(M)$ distribution
was fitted with a function:
\begin{equation}
v_2(M)=[{\rm S}(M)\cdot v_2^{\rm S}+{\rm B}(M)\cdot v_2^{\rm B}(M)]/[{\rm S}(M)+{\rm B}(M)]. 
\end{equation}
The background contribution $v_2^{\rm B}$ was 
parametrized by a linear function of $M$.
An example of the corresponding distributions and fits is shown in Fig.~\ref{fig:QCSPfit} for $\Dzero$ mesons in the interval $4<\pt<6~\gev/c$ 
with the two-particle cumulants method (a)
and $\Dstar$ mesons in the interval $2<\pt<4~\gev/c$ with the scalar
product method (b). 
The values of $v_2^{\rm S}$, hereafter indicated as $v_2\{2\}$ and $v_2\{{\rm SP}\}$,
are also reported in the figure.

Since the measured D meson yield has a feed-down contribution from B meson 
decays, the measured $v_2$ is a combination of $v_2$ of 
promptly produced and feed-down D mesons. 
In fact, the contribution of D mesons from B meson decays is enhanced by the applied 
selection criteria, because the decay vertices of the feed-down 
D mesons are, on average, more displaced from the primary vertex.
The elliptic flow of promptly produced D mesons, $v_2^{\rm prompt}$, can be 
obtained from the measured $v_2^{\rm all}$ ($v_2\{{\rm EP}\}$, $v_2\{2\}$ or $v_2\{{\rm SP}\}$) as:
\begin{equation}
\label{eq:bfeed}
v_2^{\rm prompt}=\frac{1}{f_{\rm prompt}}v_2^{\rm all} -\frac{1-f_{\rm prompt}}{f_{\rm prompt}}v_2^{\rm feed\mbox{-}down}\,,
\end{equation}
where $f_{\rm prompt}$ is the fraction of promptly produced D mesons in the 
measured raw yield and $v_2^{\rm feed\mbox{-}down}$ is the elliptic flow of D mesons
from B decays, which depends on the dynamics of beauty quarks in the medium.
These two quantities have not been measured.
According to Eq.~(\ref{eq:bfeed}), the value of $v_2^{\rm all}$ is independent of $f_{\rm prompt}$
and equal to $v_2^{\rm prompt}$,
if $v_2^{\rm feed\mbox{-}down}=v_2^{\rm prompt}$. 
The central value of the prompt D meson elliptic flow was defined
under this assumption, which removes the need to
apply the feed-down correction.
Because of the larger mass of the b quark, the $v_2$ of B mesons is expected to be lower than that of D mesons. 
Therefore, the choice of $v_2^{\rm feed\mbox{-}down}=v_2^{\rm prompt}$ as central value is the most conservative for the observation of D meson $v_2>0$.
The details of the systematic uncertainty related to this assumption are discussed in Section~\ref{sec:syst}.

\subsection{Azimuthal dependence of the nuclear modification factor}
\label{sec:raavsep}

The in-plane and out-of-plane nuclear modification factors of prompt $\rm D^0$ mesons are defined as:
\begin{equation}
\label{eq:raavsEP}
 R_{\rm AA}^{\textnormal{in (out)}}(p_{\rm T}) = \frac { 2\cdot {\rm d}N^{\textnormal{in (out)}}_{\rm AA}/ {\rm d} p_{\rm T}}
{\langle T_{\rm{AA}} \rangle \cdot {\rm d}\sigma_{pp}/{\rm d} p_{\rm T}}\,,
\end{equation} 
where ${\rm d}N^{\textnormal{in (out)}}_{\rm AA}/ {\rm d} p_{\rm T}$
are the $\rm D^0$ meson per-event yields, integrated over the two $90^\circ$-wide intervals used 
to determine $v_2$ with the event plane method. The factor 2 in
Eq.~(\ref{eq:raavsEP}) accounts for the fact that the D meson yields for Pb--Pb collisions are integrated
over half of the full azimuth.
$R_{\rm AA}^{\textnormal{in (out)}}$ was measured in the 30--50\% centrality class for $\Dzero$ mesons, which have the highest signal significance, using the yields relative to the event plane defined with TPC tracks in $0<\eta<0.8$.
The average value of the nuclear overlap function in this centrality class, $\langle T_{\rm{AA}} \rangle = 3.87\pm 0.18~{\rm mb^{-1}}$, was 
determined with the procedure described in~\cite{centrality}.

The yields of prompt $\Dzero$ mesons in the two azimuthal intervals were obtained as: 
\begin{equation}
  \label{eq:dNdpt}
  \left.\frac{{\rm d} N^{\rm D^0}}{{\rm d}\pt}\right|_{|y|<0.5}=
  \frac{1}{ \Delta y \,\Delta \pt}\frac{\left.f_{\rm prompt}(\pt)\cdot \frac{1}{2} N_{\rm raw}^{\rm D^0+\overline{D^0}}(\pt)\right|_{|y|<y_{\rm fid}}\cdot c_{\rm refl}(\pt)}{({\rm Acc}\times\epsilon)_{\rm prompt}(\pt) \cdot{\rm BR} \cdot N_{\rm events}}\,.
\end{equation}
The raw yields $N_{\rm raw}^{\rm D^0+\overline{D^0}}$ were divided by a factor of two to obtain the charge (particle and 
antiparticle) averaged yields.
The factor $c_{\rm refl}(\pt)$ was introduced to correct the raw yields for the contribution of signal candidates that are present
in the invariant mass distribution both as $\rm D^0\to K^-\pi^+$ and as $\rm \overline{D^0}\to \pi^-K^+$
(the combination with wrong mass hypothesis assignment is called `reflection').
To correct for the contribution of B meson decay feed-down, the raw yields 
were  multiplied by the prompt fraction 
$f_{\rm prompt}$, whose determination is described later in this section. 
Furthermore, they were divided by the product of prompt 
D meson acceptance and efficiency $({\rm Acc}\times\epsilon)_{\rm prompt}$, 
 normalized by the decay channel 
branching ratio ({\rm BR}), the transverse momentum ($\Delta \pt$) and rapidity ($\Delta y=2\,y_{\rm fid}$) interval widths
and the number of events ($N_{\rm events}$). 
The normalization by $\Delta y$ gives the corrected yields in one unit of rapidity $|y|<0.5$.

The $({\rm Acc}\times\epsilon)$ correction was determined, as a function of $p_{\rm T}$, 
using Monte Carlo simulations with a detailed description of the ALICE detector geometry and the GEANT3 particle transport package~\cite{geant3}.  
The simulation was tuned to reproduce the (time-dependent)
position and width of the interaction vertex distribution, as well
as the number of active electronic channels and the accuracy of the detector calibration.
The HIJING~v1.383~\cite{hijing} generator was used to simulate Pb--Pb collisions at $\sqrtsNN=2.76~\tev$ and all the produced particles were transported through the detector simulation.
Prompt and feed-down D meson signals were added using pp events from
the PYTHIA~v6.4.21~\cite{pythia} event generator with the Perugia-0
tune~\cite{perugia0}. Each simulated pp event
contained a $\rm c\overline{c}$ or $\rm b\overline{b}$ pair with D
mesons decaying into the hadronic channels of interest for the analysis. 
Out of all the particles produced in these PYTHIA pp events, only the 
heavy-flavour decay products were kept and transported through the detector 
simulation together with the particles produced by HIJING.
In order to minimize the bias on the detector occupancy,
the number of D mesons injected into each
HIJING event was adjusted according to the Pb--Pb collision centrality.

\begin{figure}[t!]
\begin{center}
\includegraphics[width=0.48\textwidth]{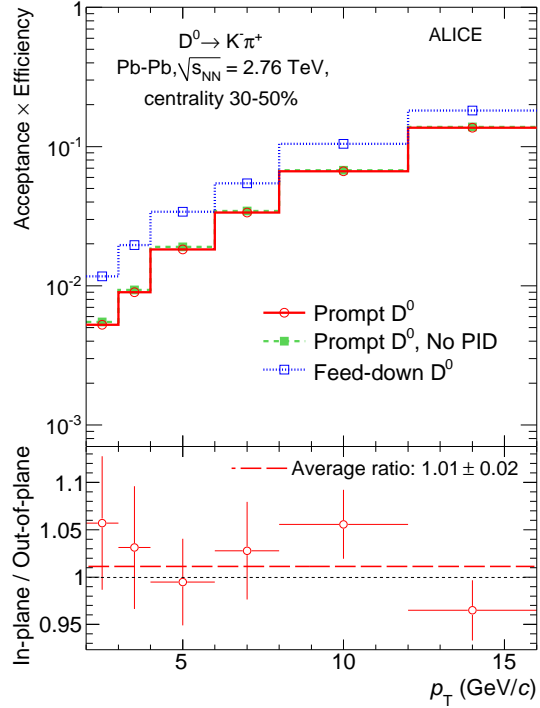}
\caption{(Color online) Product of acceptance and efficiency for $\Dzero$ mesons in Pb--Pb collisions for 30--50\% centrality class (upper panel). 
The rapidity interval is $|y|<y_{\rm fid}$ (see Section~\ref{sec:reco} for details).
The values for prompt (solid lines) and feed-down (dotted lines)
$\Dzero$ mesons are shown. Also displayed, for comparison, are the values for 
prompt $\Dzero$ mesons without PID selection (dashed lines). 
The lower panel shows the ratio of the efficiencies for prompt $\Dzero$ mesons in the in-plane and out-of-plane regions used for the analysis. This ratio was estimated using simulation samples with a difference in particle multiplicity similar to that observed in data for the two azimuthal regions.}
\label{fig:EffD0}
\end{center}
\end{figure}

The efficiencies were evaluated from 
simulated events that had the same average charged-particle multiplicity, corresponding to the same detector occupancy, as 
observed for real events in the centrality class 30--50\%. 
Figure~\ref{fig:EffD0} shows $({\rm Acc}\times\epsilon)$ for prompt and feed-down 
$\rm D^0$ mesons within the rapidity interval $|y|<y_{\rm fid}$.
The magnitude of $({\rm Acc}\times\epsilon)$ increases 
with $\pt$, starting from about 1\% and reaching about 10--15\% at high $\pt$.
Also shown in Fig.~\ref{fig:EffD0} are the $({\rm Acc}\times\epsilon)$ values for the case where no PID was applied. 
The relative difference with respect to the $({\rm Acc}\times\epsilon)$ obtained using also the PID selection is only about 5\%, thus illustrating the high efficiency of the applied PID criteria.
The $({\rm Acc}\times\epsilon)$ for D mesons from B decays is larger than 
for prompt D mesons by a factor of about 1.5,
because the decay vertices of the feed-down D mesons 
are more displaced from the primary vertex and are, therefore, more efficiently
selected by the analysis cuts. 

The possible difference in the reconstruction and selection efficiency between in-plane
and out-of-plane $\Dzero$ mesons was studied using 
simulations. This difference could arise from the variation of the particle density, and consequently of the detector occupancy, 
induced by the azimuthal anisotropy of bulk particle production. 
The difference in occupancy was estimated in data using the multiplicity of 
SPD tracklets in the two considered azimuthal intervals. Tracklets are defined as combinations of two hits in the two SPD layers that are required to point 
to the primary vertex. They can be used to measure the multiplicity of charged particles with $\pt>50~\mev/c$ and $|\eta|<1.6$.
The SPD tracklet multiplicity in the 30--50\% centrality class was found to be larger in-plane than out-of-plane by about 12\%.
In order to study the efficiency variation, two sets of simulated events with
12\% difference in average multiplicity were used.
The ratio of the two efficiencies was found to be consistent with
unity (see lower panel of Fig.~\ref{fig:EffD0}) and therefore no
correction was applied.

\begin{figure*}[t!]
\begin{center}
\includegraphics[width=0.45\textwidth]{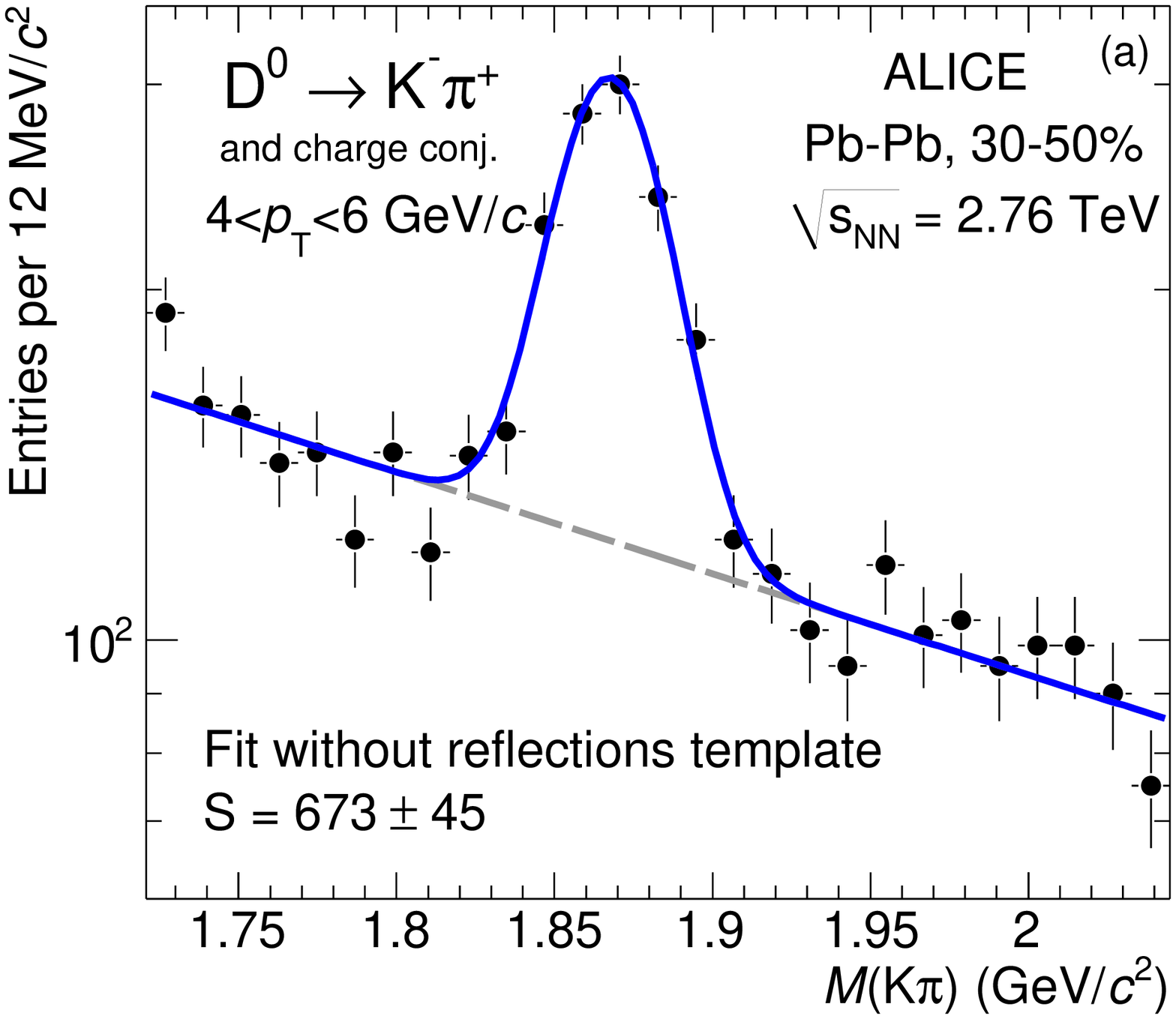}
\includegraphics[width=0.45\textwidth]{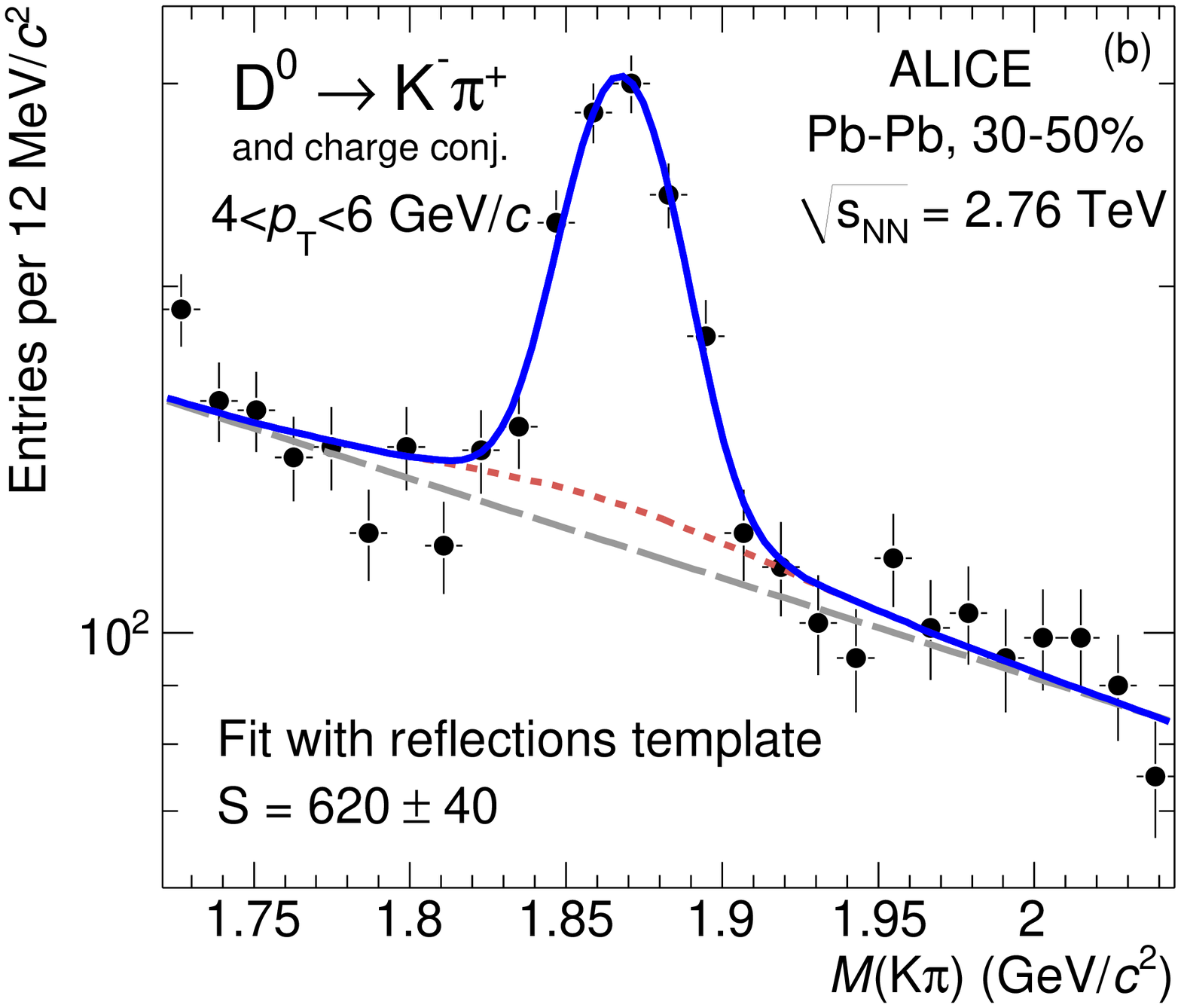}
\caption{(Color online) Invariant mass distribution of $\rm D^0$
  candidates with $4<\pt<6~\gev/c$ in the centrality class 30--50\%:
  (a) fit without template for reflections and (b) with template for reflections (dotted line). The raw yield obtained as integral of the signal Gaussian function is reported.}
\label{fig:reflections}
\end{center}
\end{figure*}

The correction factor $c_{\rm refl}$ for the contribution of reflections to the raw yield was determined by including in the invariant
mass fit procedure a template of the distribution of reflected signal candidates, which was obtained from the simulation for each $\pt$ interval.
This distribution has a centroid close to the $\rm D^0$ mass and has typical r.m.s. values of about 100~MeV/$c^2$, i.e.\,about one order of magnitude
larger than the signal invariant mass resolution. The distribution from the simulation was parametrized with the sum of two Gaussians, in order
to remove the statistical fluctuations.
In the fit with the template, the ratio of the integrals of the total distribution of reflections and of the Gaussian used for the
signal were fixed to the value obtained from the simulation. This ratio is mostly determined by the PID selection,
which limits the probability that a true $\rm K^-\pi^+$ pair can be also compatible with the $\rm \pi^-K^+$ mass hypothesis.
For the $v_2$ analysis described in the previous section, the PID selection was used only for tracks with $p<4~\gev/c$.  
Since the contribution of the reflections does not depend on the angle relative to the event plane, it is not necessary to apply 
the $c_{\rm refl}$ correction for $v_2$.
For the $\RAA$ analysis, in order to minimize the correction, the PID selection was extended to tracks with $p> 4~\gev/c$,
requiring the compatibility of the TOF and TPC signals with the expectations for kaons and pions within $3\,\sigma$. 
It was verified that this change results in a variation of $v_2$ well within the uncertainties.
The correction factor $c_{\rm refl}$ was determined as the ratio of the signal yield from the fit including the reflections template 
and the signal yield from the fit without the template.
It was computed using the sum of the in-plane and out-of-plane invariant mass distributions, 
in order to have a more precise value, and it was applied as in Eq.~(\ref{eq:dNdpt}) for both the in-plane and out-of-plane yields. 
The procedure was validated using the simulation, where the signal yield obtained from the fit with the template can be compared 
with the true signal yield.
The numerical value of $c_{\rm refl}$
ranges from 0.98 in the interval $2<\pt<3~\gev/c$ to 0.90 in the interval $4<\pt<16~\gev/c$.
Figure~\ref{fig:reflections} shows an example of the fits without (a) and with (b) template for the interval 4--6~$\gev/c$.

The fraction $f_{\rm prompt}$ of promptly produced D mesons 
in the measured raw yields
was obtained, following the procedure introduced in~\cite{aliceDRAA}, as:
\begin{equation}
  \label{eq:fcNbMethod}
\begin{split}
f_{\rm prompt} &= 1-\frac{ N^{\rm D^0\,\textnormal{feed-down}}_{\rm raw}Ê} { N^{\rm D^0}_{\rm raw} }=\\
 &= 1 -    \RAA^{\textnormal{feed-down}} \cdot    \langle \TAA \rangle 
	 \cdot 2 \cdot \left( \frac{{\rm d}^2 \sigma}{{\rm d}y \, {\rm d}\pt } \right)^{{\sf FONLL,\,EvtGen}} _{{\textnormal{ feed-down}}} \cdot  \frac{({\rm Acc}\times\epsilon)_{\textnormal{feed-down}}\cdot\Delta y \, \Delta\pt \cdot {\rm BR} \cdot N_{\rm evt}  }{ N^{\rm D^0}_{\rm raw}  } \, .
\end{split}
\end{equation}
In this expression, where the symbol of the $\pt$-dependence has been omitted for brevity, $N^{\rm D^0}_{\rm raw}$ is the measured raw yield
(corrected by the $c_{\rm refl}$ factor) 
and $N^{{\rm D^0}\,\textnormal{feed-down}}_{\rm raw}$ is the contribution of $\rm D^0$ mesons from B decays to the raw yield, estimated on the basis of 
the FONLL calculation of beauty production~\cite{fonll2012}. 
In detail, the B meson production cross section in pp collisions at $\sqrt{s}=2.76~\tev$ was folded with 
the ${\rm B\rightarrow D^0}+X$ decay kinematics using EvtGen~\cite{evtgen} and multiplied by:
the average nuclear overlap function $\langle \TAA \rangle$ in the 30--50\% centrality class,
the acceptance-times-efficiency for feed-down $\rm D^0$ mesons, and the other factors introduced in Eq.~(\ref{eq:dNdpt}).
In addition, 
the nuclear modification factor $\RAA^{\textnormal{feed-down}}$ of D mesons from B decays was accounted for.
The comparison of the $\RAA$ of prompt D
mesons~\cite{GrelliEPS} with that of $\rm J/\psi$ from B
decays~\cite{mironov} measured in the CMS experiment indicates that charmed hadrons are more suppressed than beauty hadrons. 
Therefore, it was assumed that the ratio of the nuclear modification factors 
for feed-down and prompt D mesons lies in the range $1<\RAA^{\textnormal{feed-down}}/\RAA^{\rm prompt}<3$.
The value $\RAA^{\textnormal{feed-down}}=2\cdot\RAA^{\rm prompt}$ was used to compute the correction, and the variation over the full range,
which also accounts for possible centrality and $\pt$ dependences,
was used to assign a systematic uncertainty.
The hypothesis on the nuclear modification of feed-down D mesons was changed 
with respect to the assumption used in~\cite{aliceDRAA}, based on the most recent results on the $\RAA$ of prompt D meson and non-prompt  $\rm J/\psi$  mentioned above.
As it was done for the $v_2$ measurement, the feed-down contribution was computed assuming 
$v_2^{\textnormal{feed-down}}=v_2^{\rm prompt}$. 
Therefore, the
ratio $\RAA^{\textnormal{feed-down}}/\RAA^{\rm prompt}$ is the same in-plane and out-of-plane.
The systematic uncertainty related to this assumption is discussed in Section~\ref{sec:syst}.  
For $\Dzero$ mesons, assuming $\RAA^{\textnormal{feed-down}}=2\cdot\RAA^{\rm prompt}$, the resulting $f_{\rm prompt}$
ranges from about $0.80$ in the lowest transverse momentum interval 
($2<\pt<3~\gev/c$) to about $0.75$ at high $\pt$.
 
The $\Dzero$ yields in the two azimuthal regions with respect to the event plane, obtained from Eq.~(\ref{eq:dNdpt}), were corrected for the event plane
resolution using the correction factor $R_2$ (Section~\ref{sec:methods}) and the relation given in Eq.~(\ref{eq:twobins}). For example, the correction factor for the in-plane $\RAA$ is $(1+R_2^{-1})/2+ (N^{\rm in}/N^{\rm out})\cdot (1-R_2^{-1})/2$, where $N^{\rm in\,(out)}$ is the $\rm D^0$ raw yield.
The value $R_2=0.8059\pm 0.0001$ for the
30--50\% centrality class and the typical $N^{\rm in}/N^{\rm out}$ magnitude 
result in a correction of approximately $+4\,(-6)\%$ for the in-plane (out-of-plane) yields.
 
The prompt $\Dzero$ meson production cross section in pp collisions
used in the calculation of the nuclear modification factor was 
obtained by scaling the $\pt$-differential 
cross section in $|y|<0.5$ at $\sqrt{s}=7~{\rm TeV}$, measured using a
data sample of $L_{\rm int}=5~{\rm nb}^{-1}$~\cite{Dpp7}. 
The $\pt$-dependent scaling factor was defined 
as the ratio of the cross sections obtained from FONLL calculations~\cite{fonll2012} at $\sqrt{s}=2.76$ and 7~TeV~\cite{scaling}.
The scaled $\rm D^0$ meson $\pt$-differential cross section is
consistent with that measured at $\sqrt{s}=2.76~\tev$ using a smaller statistics data sample with $L_{\rm int}=1.1~{\rm nb}^{-1}$~\cite{Dpp276},
which only covered a reduced $\pt$ interval with a statistical uncertainty of 20--25\% and was therefore not used as pp reference.
The correction for reflections was not applied for the $\rm D^0$ cross section in pp collisions. It was verified that
the resulting signal bias is smaller than 5\% ($c_{\rm refl}>0.95$), which is less than the systematic uncertainty assigned 
for the yield extraction (10--20\%~\cite{Dpp7}).

\section{Systematic uncertainties}
\label{sec:syst}

Several sources of systematic uncertainty were considered for both
$\vtwo$ and $\RAA$ analyses. The uncertainties on $v_2$
are described first. Afterwards, the systematic uncertainties affecting  $\RAA$ in-plane and
out-of-plane are discussed. The uncertainties for 
the 30--50\% centrality class
are summarized 
in Tables~\ref{tab:systv2} and~\ref{tab:systRAA}.
In the following, the quoted uncertainties are symmetric around the central value of the measurement, unless the upper and lower
parts are reported separately.

\subsection{Uncertainties on $v_2$}

One of the main sources of uncertainty originates from the 
D meson yield extraction using a fit to the invariant mass distributions. 
This uncertainty was estimated by repeating the fits under different conditions and by 
utilizing alternative methods for the yield determination.
For the $\vtwo$ analysis with the event plane method, 
the fit ranges and the functional forms for the 
combinatorial background were varied.
Polynomial and exponential functions were tried 
for $\Dzero$ and $\Dplus$ background shapes, while a threshold function multiplied by
an exponential was considered for the $\Dstar$: $a\sqrt{\Delta M-m_{\pi}}\cdot{\rm e}^{b(\Delta M-m_{\pi})}$, with 
$a$ and $b$ as free parameters.
The D meson yield was also extracted 
 by counting the entries in the invariant mass distributions after background subtraction. 
For this procedure the background was estimated with a fit to the left and right sides of the D meson invariant mass peak (side-band regions),
using the fit functions described in Section~\ref{sec:reco}.
The $\vtwo$ analysis employing
the event plane method was performed by fixing the Gaussian centroids and widths of the in-plane and out-of-plane invariant mass
distributions to the values obtained from a fit of the $\varphi$-integrated
distribution. 
The analysis was repeated with free Gaussian parameters in the fit.
The systematic uncertainty due to the yield measurement was estimated 
as the maximum variation of the $v_2$ values obtained from the described tests. It 
amounts to 10--20\% for the $\Dzero$ meson, depending on the $\pt$ and
centrality intervals, and 20--50\% for the $\Dplus$ and $\Dstar$ mesons, depending on the $\pt$ interval. 
The same procedure was applied for the two-particle correlation methods (scalar product and two-particle cumulants),
except for the bin counting method and the fixed Gaussian centroids and widths. 
Instead, the parametrization of the background $v_2^{\rm B}(M)$
was varied from a first order to a second order polynomial. 
The resulting uncertainty is in the range 15--30\%.

For the event plane method, two alternative procedures were considered
to extract $v_2$, which are not directly based on the measurement of the signal yields from the invariant mass
distribution.
These procedures use the distribution of $\cos (2\Delta\varphi)$ versus invariant mass (where $\Delta\varphi=\varphi_{\rm D}-\psi_2$)
and the relation $v_2=\langle\cos (2\Delta\varphi)\rangle$.
In the first procedure, the distribution of $\cos (2\Delta\varphi)$ is considered for
the signal region ($|M-m_{\rm D}|<3\,\sigma$) and the two side-band regions ($4<|M-m_{\rm D}|<7\,\sigma$).
The distribution of $\cos (2\Delta\varphi)$ 
for the background is obtained by averaging, bin-by-bin, the distributions of $\cos (2\Delta\varphi)$ in the two side bands. 
This background distribution is then rescaled to the integral of the background fit function in the invariant mass signal region
and  it is subtracted from the total $\cos (2\Delta\varphi)$ distribution in the signal region.
In this way, 
the distribution of  $\cos (2\Delta\varphi)$ of the signal is obtained.
Its mean value gives the D meson $v_2$.
In the second procedure, a distribution of  $\langle\cos (2\Delta\varphi)\rangle$ as a function of invariant mass is used 
for a simultaneous fit of the $v_2$ and the yield, as in the case of the two-particle correlation methods.
These two alternative procedures result in D meson $v_2$ values that are consistent with those obtained from 
the event plane method with two $\Delta\varphi$ bins. 
Therefore, no systematic uncertainty is taken for the $v_2$ extraction procedure.

The $v_2$ analysis was repeated with different sets of cuts for 
the selection of D meson candidates. 
A set of tighter and a set of looser cuts with respect to those described in
Section~\ref{sec:reco} were considered for each D meson species, thus
varying the signal yield by about 30--50\% and, consequently, the significance and the signal-to-background ratio.
The resulting $v_2$ values were found to be consistent within statistical uncertainties. Consequently, this contribution to the systematic uncertainty was neglected. 

\begin{table*}[!t]
\begin{center}
\caption{Systematic uncertainties on the measurement of $v_2$ in the 30--50\% centrality class for the interval $4<\pt<6~\gev/c$. The uncertainties are comparable in the other $\pt$ intervals.} 
\begin{footnotesize}
\begin{tabular}{|l|ccc|ccc|ccc|}
\hline
Particle & & $\Dzero$ & & & $\Dplus$ & & & $\Dstar$ & \\
          $v_2$ analysis              & $v_2\{{\rm EP}\}$ & $v_2\{{\rm SP}\}$ & $v_2\{2\}$ & $v_2\{{\rm EP}\}$ & $v_2\{{\rm SP}\}$ & $v_2\{2\}$ & $v_2\{{\rm EP}\}$ & $v_2\{{\rm SP}\}$ & $v_2\{2\}$  \\
\hline
$M$ and $v_2$ fit stability & 9\% & 10\% & 8\% & 25\% & 8\% & 17\% & 30\% & 14\% & 11\% \\
2 or 3 sub-ev. $R_2$ & 2.3\% & -- & -- & 2.3\% & -- & -- & 2.3\% & -- & -- \\
$R_2$ centrality dependence & 2\% & -- & -- & 2\% & -- & -- & 2\% & -- & -- \\
Centrality selection & -- & 10\% & 10\% & -- & 10\% & 10\% & -- & 10\% & 10\%\\
%Variation of efficiency vs. $\Psi_2$  &  & 5\% &  &   & 5\% &  &  & 5\% &  \\
\hline
Total (excl. B feed-down) & 9\% & 14\% & 13\% & 25\% & 13\% & 20\% & 30\% & 17\% & 15\%\\
\hline
B feed-down & & $^{+48}_{-0}$\% & &  & $^{+26}_{-0}$\% &  &  & $^{+26}_{-0}$\% &  \\
\hline
\end{tabular}
\end{footnotesize}
\label{tab:systv2}
\end{center}
\end{table*}

The uncertainty due to the event plane resolution 
was estimated 
with the two and three sub-event methods with an $\eta$ gap.
The three sub-events were defined using the TPC tracks and the signals in the two VZERO detectors.
The resolutions estimated with these two methods differ by 6.9\%, 2.0\% and 2.3\% 
in the 0--10\%, 10--30\% and 30--50\% centrality classes, respectively (see Fig.~\ref{fig:planes}~(b)).
A symmetric systematic uncertainty equal to the 
relative difference between $R_2$ values obtained with the two and three sub-event methods was assigned to the D meson $v_2$.

The uncertainty due to the centrality dependence of the event plane resolution 
was estimated from the difference between two ways to define the average resolution in the centrality classes used in the analysis, 
starting from the resolutions in fine centrality intervals (see Fig.~\ref{fig:planes}~(b)). Namely,
 a plain arithmetic average 
and  an average weighted with the D meson yield measured in smaller centrality classes (2.5\% wide).  
The latter was estimated using $\Dzero$ meson raw yields in wide $\pt$ intervals and the sum of the 
two $\Delta\varphi$ intervals, in order to reduce the statistical fluctuations. 
The difference between these averages was found to be about 2\%, 0.5\% and 2\% for the 
0--10\%, 10--30\% and 30--50\% centrality classes, respectively.
The resulting total uncertainties on $R_2$ amount to 
7\%, 2\% and 3\%
 for the three centrality classes.

The distribution of collision impact parameters selected in a given centrality class 
slightly depends on the pseudo-rapidity coverage of the detector used for the
centrality determination. 
The analysis was repeated using the number of tracks in the TPC 
as a centrality estimator, instead of the total signal measured in the VZERO detector. 
A relative systematic uncertainty of 10\% was assigned to the $v_2$ values measured with the scalar product and two-particle cumulant methods, on the basis of
the difference of the resulting $v_2$ values. This difference could originate from the dependence of the RFP multiplicity fluctuations on the centrality estimator.
No significant difference was observed for the event plane method when using the TPC, instead of the VZERO, for the centrality determination.

As explained at the end of section~\ref{sec:methods}, the central value of the prompt D meson $v_2$ was obtained without applying a correction 
for the feed-down from B meson decays, on the basis of the assumption $v_2^\textnormal{\rm feed-down}=v_2^{\rm prompt}$ (see Eq.~(\ref{eq:bfeed})).
The systematic uncertainty associated with this assumption 
was estimated by varying it in the interval
$0\le v_2^\textnormal{\rm feed-down}\le v_2^{\rm prompt}$.
This range covers all model predictions 
for $v_2$ of charm and beauty hadrons~\cite{gossiaux,bamps,rappv2}.
The lower limit of the variation range,
$v_2^\textnormal{\rm feed-down}=0$, gives $v_2^{\rm prompt} =v_2^{\rm all}/f_{\rm prompt}$.
The $f_{\rm prompt}$ values for each of the D meson species and each $\pt$ interval 
were obtained using FONLL calculations~\cite{fonll2012} (see section~\ref{sec:raavsep}).
Under the assumption $\RAA^\textnormal{\rm feed-down}=2\cdot\RAA^{\rm prompt}$,
the $f_{\rm prompt}$ values change from 0.8 to 0.75 (0.85 to 0.8) from low to high $\pt$ for $\rm D^0$ 
($\rm D^+$ and $\rm D^{*+}$) mesons
(the feed-down contribution is larger for $\rm D^0$ mesons because of the stronger constraint 
on the separation between the secondary and the primary vertex).  
A set of $f_{\rm prompt}$ values was computed 
by varying the heavy-quark masses
and the perturbative scales in the FONLL calculation 
as prescribed in~\cite{fonll2012},
and the ratio $\RAA^\textnormal{\rm feed-down}/\RAA^{\rm prompt}$ in the range 
$1<\RAA^\textnormal{\rm feed-down}/\RAA^{\rm prompt}<3$.
The smallest value of $f_{\rm prompt}$ was used to assign 
the uncertainty related to the B feed-down contribution to the elliptic flow of 
prompt D mesons.
The maximum relative uncertainty is about $^{+45}_{-0}\%$.

\subsection{Uncertainties on $\RAA$}

For the analysis of the $\Dzero$ meson $\RAA$ in-plane and out-of-plane, the same sources
of systematic uncertainty as for the $\vtwo$ measurement with the event plane method were considered. 
Additional systematic uncertainties, which are specific to the $\RAA$ measurement, stem from the tracking, selection and particle identification efficiencies,
and from the uncertainty 
of the proton--proton reference yield. 
The evaluation of these uncertainties is similar as in~\cite{aliceDRAA} and it is described in the following.

In order to reduce the statistical fluctuations,
the uncertainty of the $\Dzero$ yield extraction was estimated using the $\varphi$-integrated invariant mass distributions. 
The fit procedure was varied, as described for the $v_2$ analysis.
The resulting uncertainty is 7\% for $2<\pt<8~\gev/c$ and 10\% for $8<\pt<16~\gev/c$.
The systematic uncertainty on the correction factor for signal reflections, $c_{\rm refl}$, was estimated by changing 
by $\pm 50\%$ the ratio of the integral of the reflections over the integral of the signal obtained from the simulation
and used in the invariant mass fit with the reflections template. 
In addition, the shape of reflections invariant mass distribution 
template was varied using a polynomial parametrization of the distribution 
from the simulation,
 instead of a double-Gaussian parametrization.
These variations resulted in an uncertainty of 1--2\% for $2<\pt<4~\gev/c$
and of 5\% for $4<\pt<16~\gev/c$ on the $c_{\rm refl}$ factor.

The systematic uncertainty of the tracking efficiency
was estimated by comparing the probability to match the TPC tracks extrapolated to the ITS hits in
 data and simulation, and by varying the track quality selection
 criteria (for example, the minimum number of associated hits in the TPC and in the ITS and maximum $\chi^2$/ndf of the momentum fit).
The efficiency of the track matching and the association of hits in the 
silicon pixel layers was found to be described by the simulation with maximal deviations on the 
level of 5\% in the $\pt$ range relevant for this analysis 
(0.5--15~GeV/$c$). 
The effect of misassociating ITS hits to tracks was studied using simulations. It was found that the fraction of D mesons with at least one decay track with a wrong hit associated increases with centrality, due to the higher detector occupancy, and vanishes at high $\pt$, where the track extrapolation between ITS layers is more precise. In the centrality class 30--50\%, this fraction is about 2\% 
in the transverse momentum interval $2 < \pt < 16~\gev/c$. It was verified that the signal selection efficiencies are the same for D mesons with and without wrong hit associations. 
The total systematic uncertainty of the track reconstruction procedure amounts to 5\% for single tracks, 
which results in a 10\% uncertainty for $\Dzero$ mesons (two-track final state).

The uncertainty of the correction for the selection on the decay topology was evaluated by repeating the analysis with different sets of cuts and was defined as the variation 
of the resulting corrected yields with respect to the value corresponding to the baseline cuts. 
This resulted in a variation up to 10\% in the $\pt$ intervals used in the analysis.
The analysis was repeated without applying the PID selection and the resulting corrected yields were found to be 
consistent within 5\% with those obtained with the PID selection. 
Therefore, a systematic uncertainty of 5\% was assigned for the PID efficiency correction in the simulation.

The uncertainty of the efficiencies arising from the difference between the real and simulated D meson momentum distributions depends on the width of the $\pt$ intervals and on the variation of the efficiencies within them. This uncertainty includes also the effect of the $\pt$ dependence of the nuclear modification factor.
The mean efficiency in a given $\pt$ interval was computed by re-weighting 
the simulated $\Dzero$ meson yield according to the $\pt$ distribution measured for 
$\Dzero$ mesons in central Pb--Pb collisions~\cite{aliceDRAA}. 
The systematic uncertainty was defined as the difference with respect to the 
efficiency computed using the $\pt$ distribution  from a FONLL calculation~\cite{fonll2012} multiplied by the $\RAA$ value 
from one of the models~\cite{bamps} that closely describe the central value of the measurement (see Section~\ref{sec:models}). 
This uncertainty is of 2\% in the interval $2<\pt<3~\gev/c$, where the efficiency increases steeply with $\pt$, and 
below 1\% for $\pt>3~\gev/c$.

The uncertainty of 3\% on the event plane resolution correction factor $R_2$ in the 30--50\% centrality class 
was propagated to the $\RAA$ observables, resulting in an uncertainty in the range 0.5--2\%, depending on the $\pt$ interval.

The systematic uncertainty due to the subtraction of feed-down D mesons from B meson
decays was estimated following the procedure described in~\cite{aliceDRAA}.
The contribution of the 
uncertainties inherent in the FONLL perturbative calculation was included by varying the heavy-quark masses and the 
factorization and renormalization scales in the ranges proposed
in~\cite{fonll2012}.
This contribution partly cancels in the $\RAA$ ratio, because these variations 
are done simultaneously for the Pb--Pb yield and for the pp reference cross section.
The uncertainty introduced by the hypothesis on the value of the feed-down D meson $\RAA$
was estimated from the variation $1<\RAA^\textnormal{\rm feed-down}/\RAA^{\rm prompt}<3$.
The total uncertainty due to the feed-down correction,
which is common to the in-plane and out-of-plane $\RAA$,  
ranges between $^{+9}_{-13}$\% at low $\pt$ and $^{+14}_{-12}$\% at high $\pt$.
The hypothesis on the value of $v_2$ for D mesons from B decays, 
that was varied in the range $0\le v_2^\textnormal{\rm feed-down}\le v_2^{\rm prompt}$, introduces an additional 
contribution to the systematic uncertainty, which is anti-correlated between $\RAA^\textnormal{in-plane}$
and $\RAA^\textnormal{out-of-plane}$. This uncertainty is typically of $^{+5}_{-0}\%$ for in-plane and $^{+0}_{-5}\%$ for out-of-plane.

The uncertainty of the pp reference used for the calculation of $\RAA$ has two contributions. 
The first is due to the systematic uncertainty of the measured 
$\Dzero$ meson $\pt$-differential yield at $\sqrt{s}=7~\tev$ and it is 
about 17\%, approximately constant with $\pt$~\cite{Dpp7}. 
The second contribution is due to the scaling to $\sqrt{s}=2.76~\tev$.
It ranges from $^{+31}_{-10}\%$ at low $\pt$ to about 5\% at high 
$\pt$~\cite{aliceDRAA}.

\begin{table*}[!t]
\begin{center}
\caption{Systematic uncertainties on the measurement of the $\Dzero$ meson $\RAA$ in-plane and out-of-plane in the 30--50\% centrality class for two $\pt$ intervals.
The uncertainties are grouped according to the type of correlation between the in-plane and out-of-plane cases.}
\begin{footnotesize}
\begin{tabular}{|l|ccc|}
\hline
%Particle & & $\Dzero$ & \\
$\pt$ interval (GeV/$c$) & 2--3 & & 12--16 \\
\hline
Uncorrelated uncertainties & & & \\
~~~~~~Yield extraction & 7\% & & 10\% \\
& & & \\
Total uncorrelated & 7\% & & 10\% \\
\hline
Correlated uncertainties & & & \\
~~~~~~Correction for reflections & 1\% & & 5\% \\
~~~~~~Tracking efficiency & 10\% & & 10\% \\
~~~~~~Cut efficiency & 10\% & & 10\% \\
~~~~~~PID efficiency & 5\% & & 5\% \\
~~~~~~$\Dzero$ $\pt$ distribution in MC & 2\% & & 0 \\
~~~~~~pp reference & $^{+20}_{-35}\%$ & & 18\% \\
~~~~~~~~~~~Data syst. & 17\% & & 17\% \\
~~~~~~~~~~~$\sqrt{s}$ scaling & $^{+10}_{-31}\%$ & & $^{+5}_{-6}\%$ \\
~~~~~~B feed-down yield & $^{+9}_{-13}\%$ & & $^{+14}_{-12}\%$ \\
& & & \\
Total correlated & $^{+22}_{-37}$\% & & $^{+28}_{-27}$\% \\
\hline
Normalization uncertainties &  &  &  \\
~~~~~~pp cross section norm. &  & 3.5\% &  \\
~~~~~~$\av{\TAA}$ &  & 4.7\% &  \\
~~~~~~Centrality class definition &  & 2\% &  \\
& & & \\
Total normalization &  & 6.2\% &  \\
\hline
Anti-correlated uncertainties & & & \\
%~~~~~~Variation of efficiency vs. $\Psi_2$ &  & 3\% &  \\
~~~~~~Uncertainty on $R_2$ & 0.5\% & & 0.5\% \\
~~~~~~B feed-down $v_2$ & in: $^{+4}_{-0}\%$; out: $^{+0}_{-6}\%$& & in: $^{+7}_{-0}\%$; out: $^{+0}_{-5}\%$ \\
& & & \\
Total anti-correlated & in: $^{+4}_{-0.5}\%$; out: $^{+0.5}_{-6}\%$ & & in: $^{+7}_{-0.5}\%$; out: $^{+0.5}_{-5}\%$ \\
\hline
\end{tabular}
\end{footnotesize}
\label{tab:systRAA}
\end{center}
\end{table*}

The uncertainties on the pp cross section normalization (3.5\%)~\cite{Dpp7} and the 
average nuclear overlap function $\langle\TAA \rangle$ (4.7\% for the class 30--50\%) were also included.
The contribution due to the 1.1\% relative uncertainty on the fraction of the
hadronic cross section used in the Glauber fit to determine the centrality 
classes~\cite{centrality} was obtained by estimating the variation of the D meson 
$\d N/\d\pt$ 
when the limits of the centrality classes are shifted by $\pm$1.1\% 
(e.g.\,instead of 30--50\%, 30.3--50.6\% and 29.7--49.5\%)~\cite{aliceDRAA}. 
The resulting uncertainty, common to all $\pt$ intervals, is 
2\% for the 30--50\% centrality class. The total normalization uncertainty, computed taking the quadratic sum of these three contributions,
is 6.2\%.

The systematic uncertainties of $\RAA$ were grouped in three
categories, depending on their correlation between the in-plane
and out-of-plane measurements.
The uncorrelated systematic uncertainties affect the two $\RAA$ 
independently; this category includes
only the yield extraction uncertainty. 
The correlated systematic uncertainties affect the two 
$\RAA$ in the same way and do not affect their relative difference. 
The uncertainties on the correction efficiencies (for track reconstruction, selection cuts, 
 particle identification and $\Dzero$ $\pt$ distribution in the simulation) and on the correction for reflections, as well as those on 
the pp reference, the variation of pQCD scales and the $\RAA^\textnormal{\rm feed-down}$ hypothesis used 
for the feed-down subtraction are included in this category.
Another correlated uncertainty is due to the normalization ($\av{\TAA}$ and centrality class definition), which is quoted separately.
The anti-correlated systematic uncertainties could shift the two $\RAA$ in opposite directions,
affecting their difference. 
This category includes 
the contribution from the unknown azimuthal anisotropy of feed-down D mesons
(variation of $v_2^\textnormal{\rm feed-down}$)
and the contribution from the event plane resolution correction factor. 
Within each category, the uncertainties from different sources were added in quadrature.

\section{Results}
\label{sec:results}

\subsection{Elliptic flow}
\label{sec:v2}

\begin{figure*}[!t]
\begin{center}
\includegraphics[width=\textwidth]{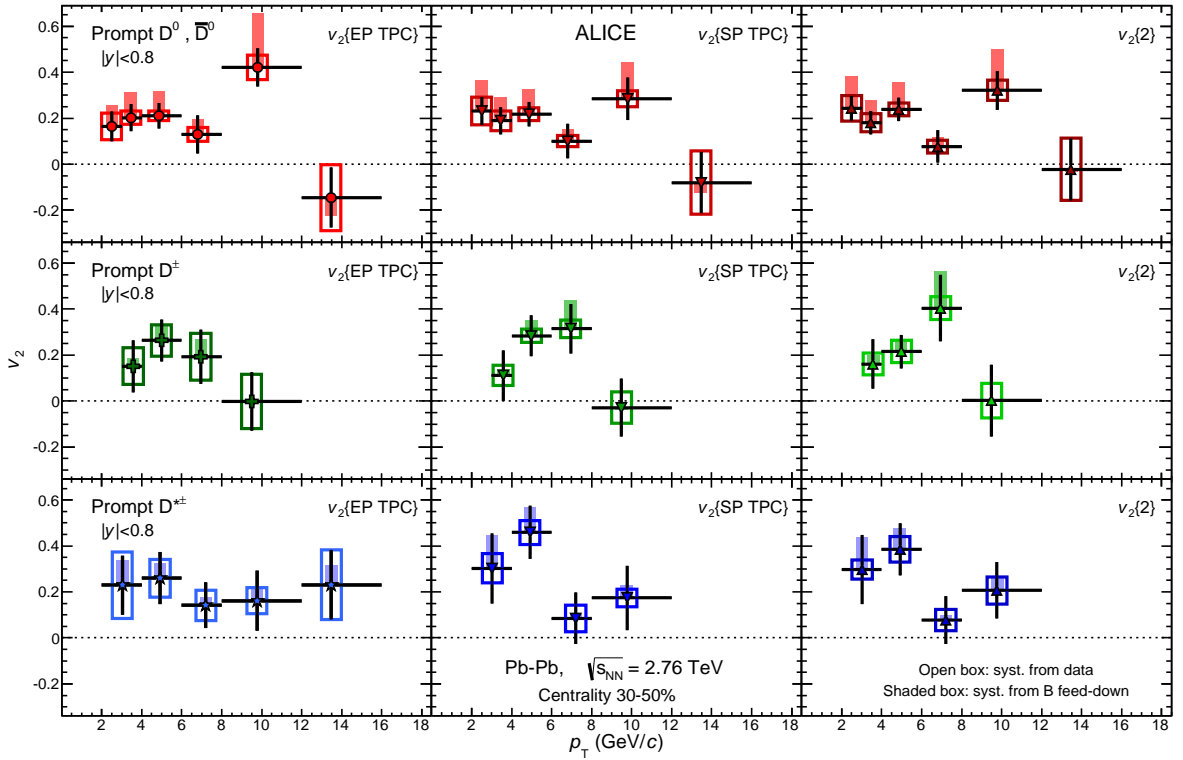}
\caption{(Color online) $v_2$ as a function of $\pt$ in the 30--50\% centrality class, for $\rm D^0$, $\rm D^+$ and $\rm D^{*+}$ mesons (rows) with the event plane (from Ref.~\cite{Dv2Letter}), scalar product and two-particle cumulant methods (columns). For the first method, the event plane
was estimated with TPC tracks in $0<\eta<0.8$; for the other methods, TPC tracks in $-0.8<\eta<0.8$ were used as RFP. The symbols are positioned at the
average $\pt$ measured within each interval.}
\label{fig:v2_3x3}
\end{center}
\end{figure*}

The elliptic flow $v_2$ measured with the event plane method is shown as a function of $\pt$ in 
the left column of Fig.~\ref{fig:v2_3x3} for $\Dzero$, $\Dplus$ and $\Dstar$ mesons
in the 30--50\% centrality class. The event plane was estimated from TPC tracks
in the range $0<\eta<0.8$.
The symbols are positioned horizontally at the average $\pt$ of 
reconstructed D mesons. 
This value was determined as the average of the $\pt$ distribution of 
candidates in the signal invariant mass region, after subtracting the 
contribution of the background candidates, which was estimated from the side bands.
This average $\pt$ of the reconstructed D mesons is larger 
than that of the produced D mesons, because the efficiency increases with increasing $\pt$ (see Fig.~\ref{fig:EffD0}).
The vertical error bars represent the statistical uncertainty, the open boxes are the
systematic uncertainties from the anisotropy determination and the event plane
resolution, and the filled boxes are the uncertainties due to the B feed-down 
contribution.
The elliptic flow of the three D meson species is consistent within 
statistical uncertainties and ranges between 0.1 and 0.3 in the interval $2<\pt<8~\gev/c$. For $\pt>12~\gev/c$, $v_2$ is consistent with zero
within the large statistical uncertainties.
The central and right-most panels of the same figure show the $v_2$ results obtained with the
scalar product and two-particle cumulant methods, respectively.
The results from the three methods are
consistent within statistical uncertainties for the three meson species.

\begin{figure*}[!t]
\begin{center}
\includegraphics[width=\textwidth]{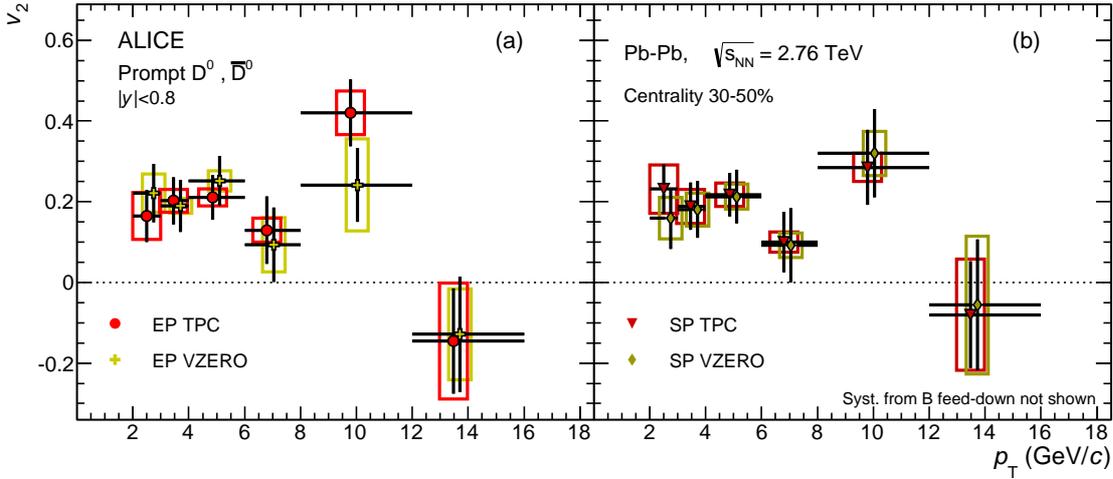}
\caption{(Color online) $\rm D^0$ meson $v_2$ as a function of $\pt$ in the 30--50\% centrality class, with the reference particles from the TPC or from the VZERO detectors ($-3.7<\eta<-1.7$ and $2.8<\eta<5.1$). (a) Event plane method. (b) Scalar product method. For visibility, the symbols for the VZERO case are slightly displaced horizontally.}
\label{fig:v2D0TPCVZERO}
\end{center}
\end{figure*}

Figure~\ref{fig:v2D0TPCVZERO} shows the $v_2$ of the $\Dzero$ mesons measured with the event plane (a)
and scalar product (b) methods using reference particles from the TPC detector (i.e.\,in a $\eta$ range that overlaps with the D meson acceptance)
or from the VZERO detectors at $-3.7<\eta<-1.7$ and $2.8<\eta<5.1$ (i.e.\,with a large $\eta$ gap with respect to the D mesons).
The agreement between the results with and without $\eta$ gap indicates that the bias due to non-flow 
correlations is within the statistical precision of the measurement.

For the 30--50\% centrality class an average $v_2$ of $\Dzero$, $\Dplus$ and $\Dstar$  
was already computed in~\cite{Dv2Letter} from the event plane method results, using the statistical uncertainties as weights. 
The resulting D meson $v_2$
has a value 
$0.204\pm 0.030\,{\rm (stat)}\,\pm  0.020\,{\rm (syst)}\,_{-0}^{+0.092}\,\textnormal{\rm (B feed-down)}$, averaged over the $\pt$ intervals 
2--3, 3--4, \mbox{4--6~$\gev/c$}. 
This value is larger than zero with a significance, calculated from the combined statistical and systematic uncertainties, 
of $5.7\,\sigma$.

\begin{figure*}[!t]
\begin{center}
\includegraphics[width=\textwidth]{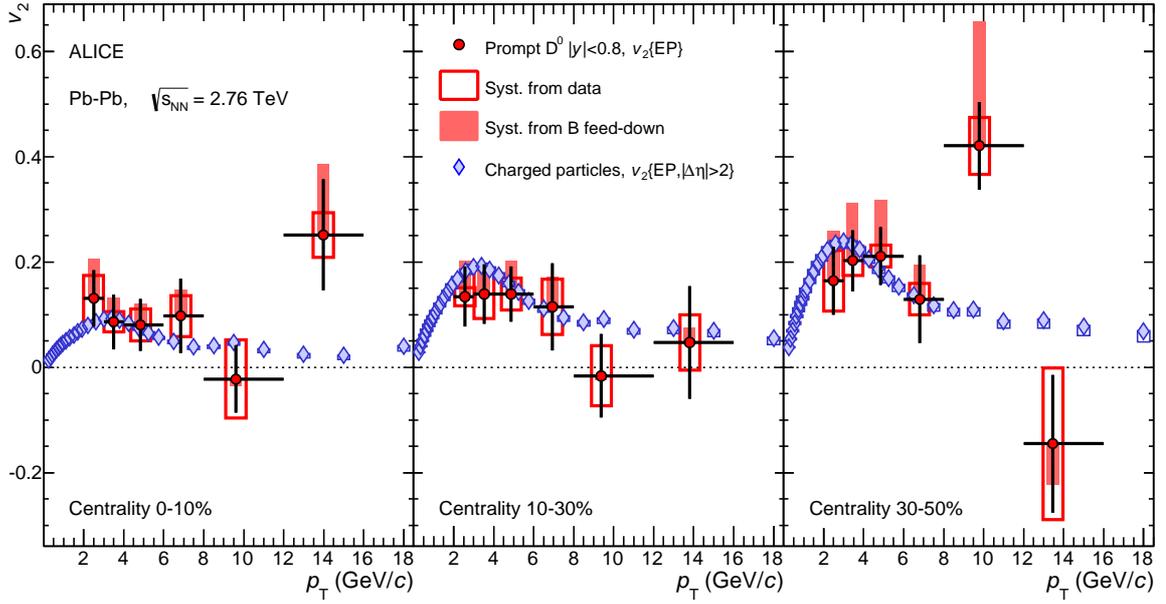}
\caption{(Color online) Comparison of prompt $\rm D^0$ meson and charged-particle $v_2$~\cite{ALICEhighptv2} in three centrality classes as a function of $\pt$.
Both measurements are done with the event plane method. For charged particles a gap of two $\eta$ units is used.}
\label{fig:v2D0cmpcharged}
\end{center}
\end{figure*}

Figure~\ref{fig:v2D0cmpcharged} shows the $\Dzero$ meson $v_2$ in the three centrality classes 0--10\%, 10--30\% and 30--50\% 
as a function of $\pt$. 
The $\Dzero$ meson $v_2$ is compared with that of charged particles~\cite{ALICEhighptv2}, 
for the same centrality classes. D meson and charged particle results are obtained with the event plane method using TPC  and the VZERO detectors, respectively.
The magnitude of $v_2$ is similar for charmed hadrons and light-flavour hadrons, which dominate the charged-particle sample.

\begin{figure}[!t]
\begin{center}
\includegraphics[width=0.48\textwidth]{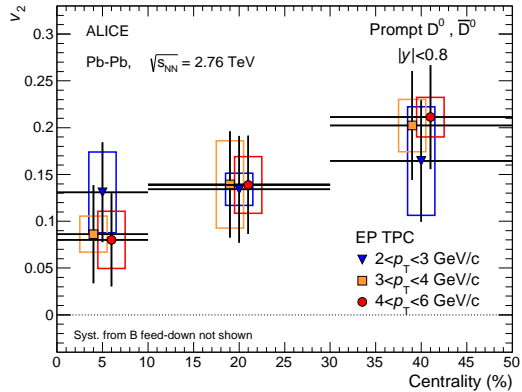}
\caption{(Color online) $\rm D^0$ meson $v_2$ with event plane method in three $\pt$ intervals as a function of centrality. For visibility, the points are displaced horizontally
for two of the $\pt$ intervals.}
\label{fig:v2D0centrality}
\end{center}
\end{figure}

The centrality dependence of the $\Dzero$ elliptic flow is shown in 
Fig.~\ref{fig:v2D0centrality} for three transverse momentum intervals 
in the range $2<\pt<6~\gev/c$.
A decreasing trend of $v_2$ towards more central 
collisions is observed, as expected because of the decreasing initial geometrical anisotropy.

%%%
%%%
%%%
\subsection{Nuclear modification factor in and out of the event plane}
\label{sec:RAAvsEP}

\begin{figure}[!t]
\begin{center}
\includegraphics[width=0.48\textwidth]{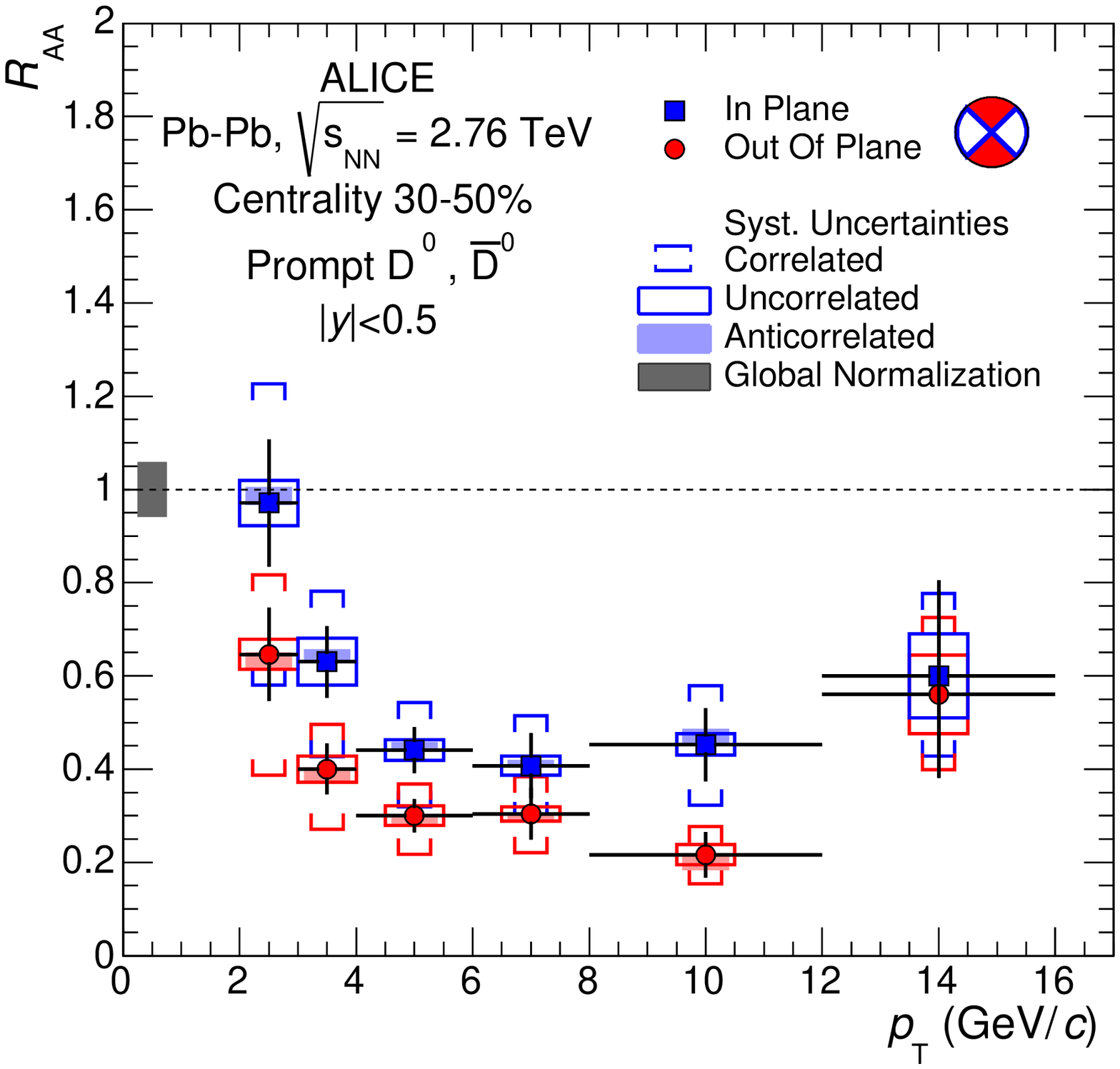}
\caption{(Color online) Nuclear modification factor $\RAA$ of $\Dzero$ mesons in the 30--50\% centrality class in two $90^\circ$-wide azimuthal intervals centred on the in-plane and on the out-of-plane directions. The correlated, uncorrelated, and anti-correlated contributions to the systematic uncertainty are shown separately.
}
\label{fig:RAAvsEP}
\end{center}
\end{figure}

The nuclear modification factors of $\Dzero$ mesons in the 30--50\% centrality class are shown in Fig.~\ref{fig:RAAvsEP} 
for the in-plane and out-of-plane directions with respect to the 
event plane. The event plane was estimated with TPC tracks in $0<\eta<0.8$. 
The error bars represent the statistical uncertainties, which are to a large extent independent for the two azimuthal intervals, since they are dominated by the statistical uncertainties of the Pb--Pb data.
The uncorrelated (empty boxes), correlated (brackets) and anti-correlated (shaded boxes)  systematic uncertainties are shown separately.
The normalization uncertainty, shown as a box at $\RAA=1$, is common to both measurements. 

A large suppression is observed in both directions with respect to the event plane for $\pt>4~\gev/c$. At lower transverse momentum, the suppression appears
to be reduced, especially in the in-plane direction, where $\RAA$ reaches unity at a $\pt$ of 2--3~$\gev/c$. 
Overall, a stronger suppression in the out-of-plane direction is observed. The ordering $\RAA^\textnormal{out-of-plane}<\RAA^\textnormal{in-plane}$
is equivalent to the observation of $v_2>0$ (as shown in the top-left panel of Fig.~\ref{fig:v2_3x3}), since Eq.~(\ref{eq:twobins}) can be expressed also as 
\begin{equation}
v_2 =\frac{\pi}{4}\frac{\RAA^{\textnormal{in-plane}}-\RAA^{\textnormal{out-of-plane}}}{\RAA^{\textnormal{in-plane}}+\RAA^{\textnormal{out-of-plane}}}\,.
\end{equation}

\section{Comparison with model calculations}
\label{sec:models}

\begin{figure*}[t!]
\begin{center}
\includegraphics[width=0.49\textwidth]{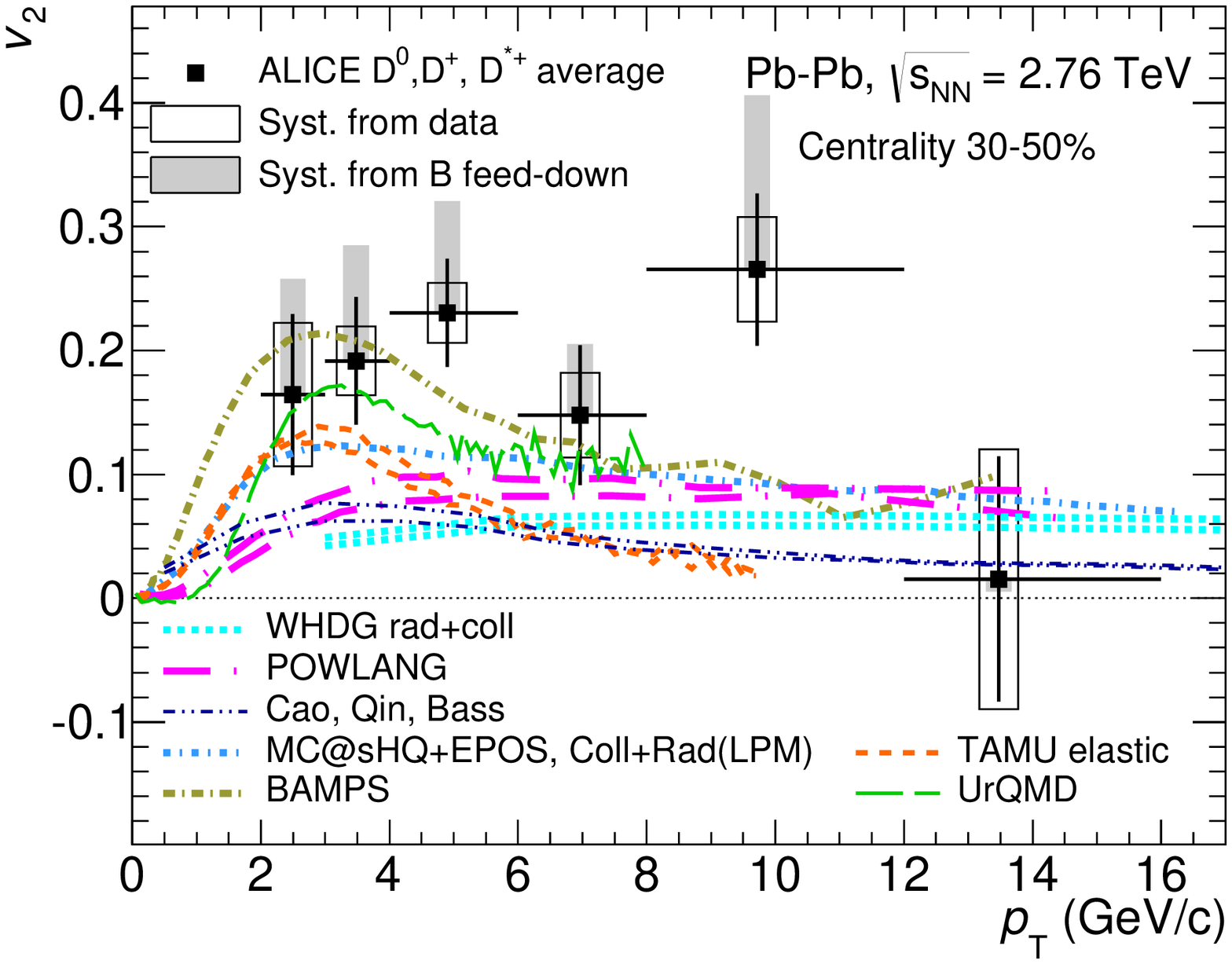}
\includegraphics[width=0.49\textwidth]{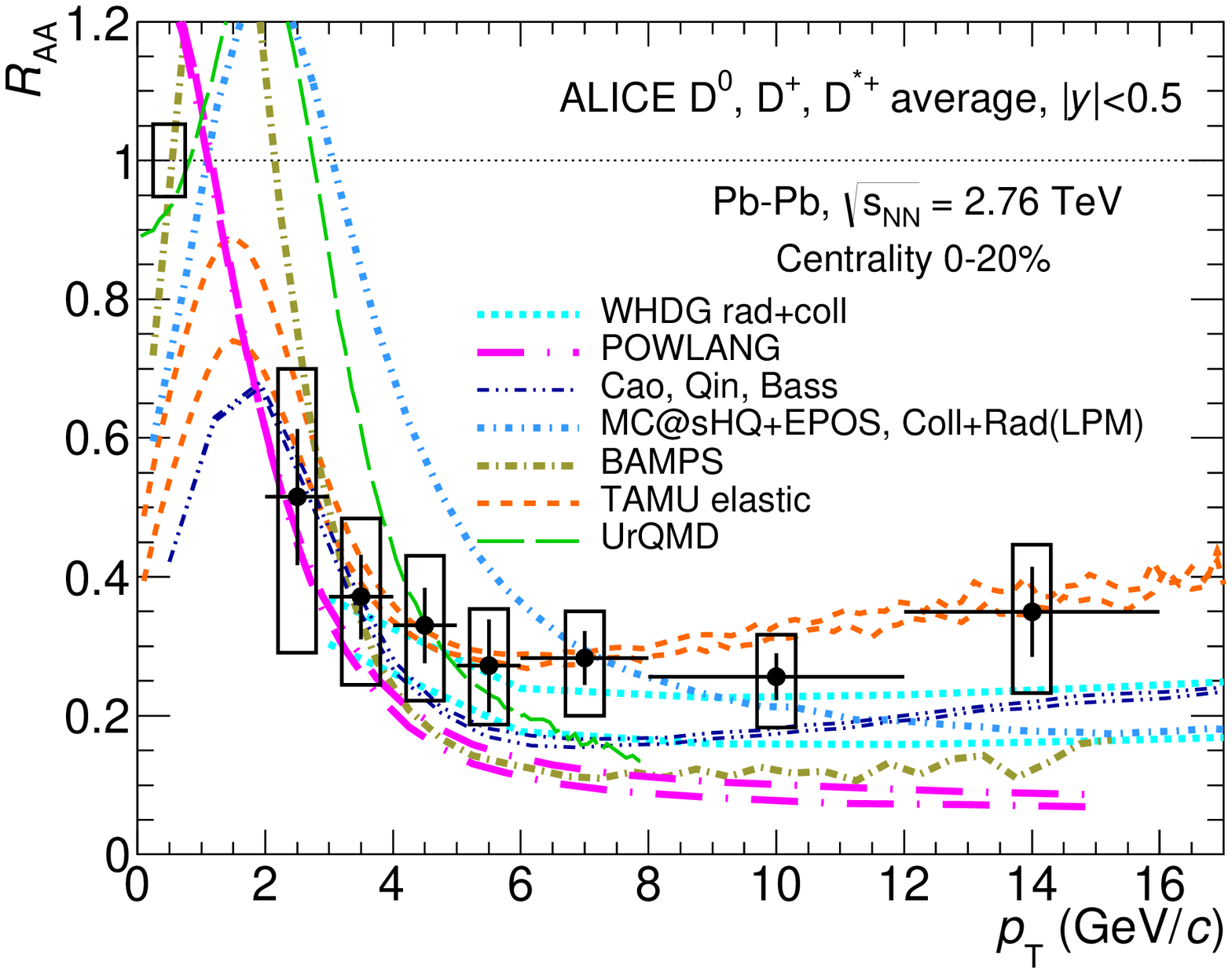}
\includegraphics[width=0.85\textwidth]{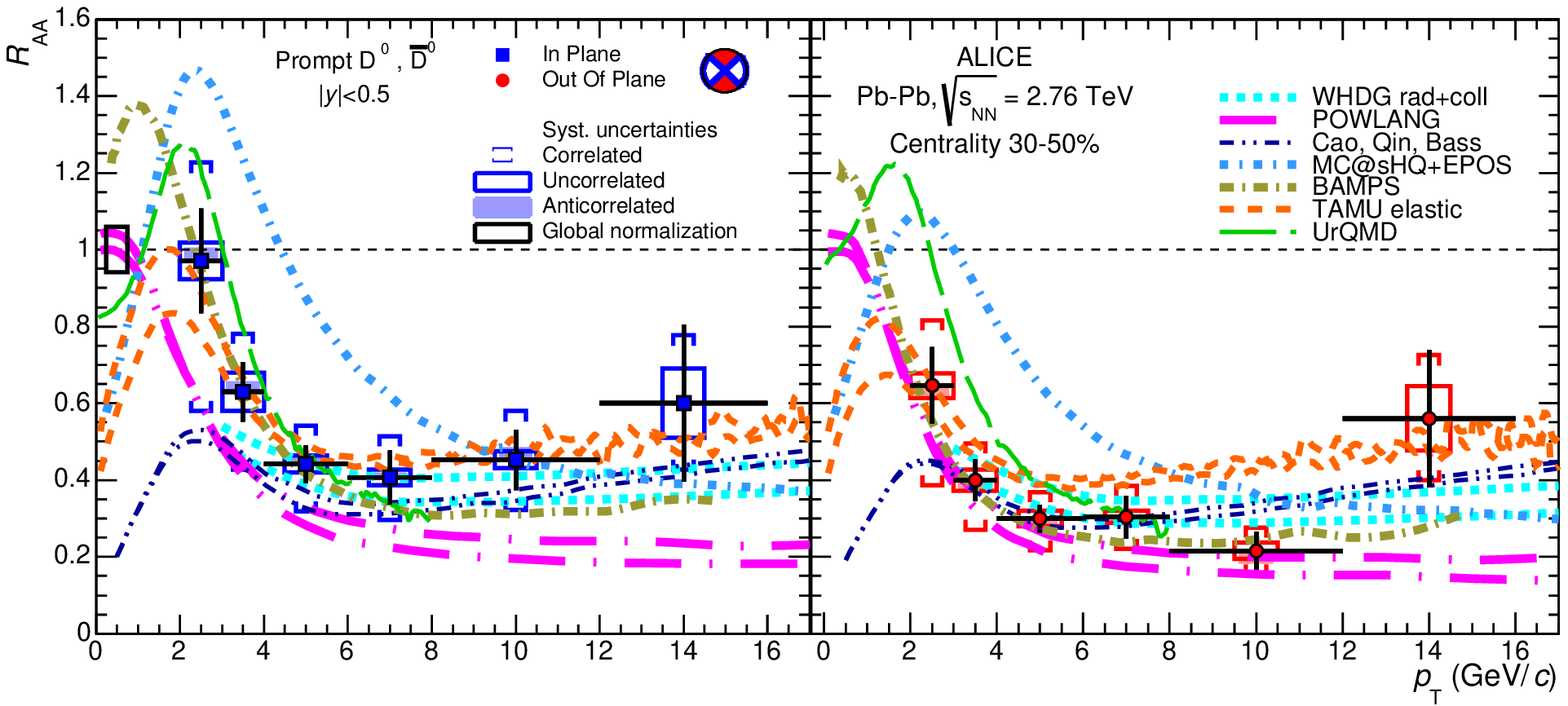}
\caption{(Color online) Model comparisons for average D meson $v_2$ in the 30--50\% centrality class (upper-left), average D meson $\RAA$ in
the 0--20\% centrality class (upper-right)~\cite{aliceDRAA}, $\Dzero$ $\RAA$ in-plane and out-of-plane in the 30--50\% centrality class (lower panels).
The seven model calculations are described in the text: WHDG rad+coll~\cite{whdg2011}, POWLANG~\cite{beraudo}, Cao, Qin, Bass~\cite{bass}, MC@sHQ+EPOS, Coll+Rad(LPM)~\cite{gossiauxEPOS}, BAMPS~\cite{bamps}, TAMU elastic~\cite{rapp2014}, UrQMD~\cite{lang}. 
The models WHDG rad+coll, POWLANG, TAMU elastic and UrQMD are shown by two lines that represent their uncertainty.
}
\label{fig:models}
\end{center}
\end{figure*}

A number of theoretical model calculations are available for the elliptic flow coefficient $v_2$ and the nuclear modification factor
$\RAA$ of heavy-flavour hadrons. 
Figure~\ref{fig:models} shows 
a comprehensive comparison of these models to measurements of the $\RAA$ of $\Dzero$ mesons in-plane and out-of-plane in the 30--50\% centrality class, 
of the average $\RAA$ of $\Dzero$, $\Dplus$ and $\Dstar$ in the  0--20\% centrality class~\cite{aliceDRAA}, and of the $v_2$
averaged over the D meson species in the centrality class 30--50\%~\cite{Dv2Letter}.

The following models are considered and compared to data:
\begin{itemize}
\item[I] WHDG~\cite{whdg2011}. This is a perturbative QCD calculation of parton energy loss, 
  including both radiative (DGLV~\cite{dglv}) and collisional processes. 
  A realistic collision geometry based on the Glauber model~\cite{glauber} is used, without hydrodynamical expansion, so that the anisotropy
  results only from path-length dependent energy loss.
  Hadronization is performed using vacuum fragmentation functions.
 The medium density is constrained on the basis of the $\pi^0$ $\RAA$ in central collisions 
 at $\sqrtsNN=200~\gev$ and scaled to LHC energy according to the increase of the charged-particle multiplicity. 
The model describes well the D meson $\RAA$ in the centrality interval 0--20\% (slightly overestimating the suppression, as it 
does also for charged particles~\cite{aliceDRAA}), and gives an almost $\pt$-independent $v_2\approx 0.06$, which 
is smaller than the measured values in the range $2<\pt<6~\gev/c$. Consequently, the difference between the
in-plane and out-of-plane $\RAA$ suppression is underestimated: the model describes well the out-of-plane $\RAA$ and
lies below the in-plane $\RAA$. 
\item[II] MC@sHQ+EPOS, Coll+Rad(LPM)~\cite{gossiauxEPOS}. This pQCD model includes 
   collisional and radiative (with Landau-Pomeranchuk-Migdal correction~\cite{bsz_lpm}) energy loss mechanisms for heavy quarks with running 
   strong coupling constant. 
   The medium fluid dynamical expansion is based on the EPOS model~\cite{EPOS}.
   A component of recombination of heavy quarks with light-flavour quarks from the QGP is also incorporated in the model.
This model yields a substantial anisotropy ($v_2\approx 0.12$--0.08 from low to high $\pt$), which is close to that observed in data.
The nuclear modification factor is substantially overestimated below $\pt\approx 5~\gev/c$ and correctly described at higher $\pt$.
\item[III] TAMU elastic~\cite{rapp2014}.  This is a heavy-flavour transport model based on collisional, elastic 
processes only. 
The heavy-quark transport coefficient is calculated within a non-perturbative $T$-matrix approach,
where the interactions proceed via resonance formation that transfers momentum from the 
heavy quarks to the medium constituents. 
The model includes hydrodynamic medium evolution, constrained by light-flavour hadron spectra and elliptic flow data,
and a component of recombination of heavy quarks with light-flavour quarks from the QGP.
Diffusion of heavy-flavour hadrons in the hadronic phase is also included. 
The model provides a good description of the observed suppression of D mesons over the entire $\pt$ range.  The maximum anisotropy, $v_2$ of about 0.13 at $2<\pt<4~\gev/c$, is close to that observed in the data. Towards larger $\pt$, the model tends to underestimate $v_2$, as well as the difference of the in-plane and out-of-plane $\RAA$.  
 \item[IV] POWLANG~\cite{beraudo}. This transport model is
 based on collisional processes treated within the framework of Langevin dynamics,
  within an expanding deconfined medium described by relativistic viscous hydrodynamics. 
  The transport coefficients entering into the relativistic Langevin equation are evaluated 
  by matching the hard-thermal-loop calculation of soft collisions with a perturbative QCD calculation for hard scatterings. 
  Hadronization is implemented via vacuum fragmentation functions.
  This model overestimates the high-$\pt$ suppression, it yields a value for $v_2$ significantly smaller than observed in data
  and also underestimates the difference between the in-plane and out-of-plane suppression. 
 \item[V] BAMPS~\cite{bamps}. This  partonic transport model is based on the
  Boltzmann approach to multi-parton scattering.
  Heavy quarks interact with the medium via collisional processes computed with running strong coupling constant.
  Hadronization is performed using vacuum fragmentation functions.
  The lack of radiative processes is accounted for by scaling the binary cross section with a correction factor, which is tuned to describe the heavy-flavour decay electron elliptic flow and nuclear modification factor at RHIC.
When applied to calculations for LHC energy, this correction factor results in an underestimation of the D meson $\RAA$ for $\pt>5~\gev/c$ 
and a large azimuthal anisotropy, with $v_2$ values up to 0.20, similar to those observed in the data. The nuclear modification factors in-plane and out-of-plane
are well described up to $5~\gev/c$, while for higher $\pt$ the in-plane $\RAA$ is underestimated.
\item[VI] UrQMD~\cite{lang}. The Langevin approach for the transport of heavy quarks is in this case implemented within the UrQMD model~\cite{urqmd}.
This model includes a realistic description of the medium evolution by combining hadronic transport and ideal hydrodynamics. The transport of heavy quarks is calculated on the basis of a resonance model with a decoupling temperature of 130~MeV. Hadronization via quark coalescence is included. The calculation parameters 
are tuned to reproduce the heavy-flavour measurements at RHIC ($\sqrtsNN=200~\gev$) and kept unchanged for calculations at the LHC energy. The model describes the measured D meson $v_2$, as well as $\RAA$ in the interval $4<\pt<8~\gev/c$, but it fails to reproduce the 
significant suppression measured for $\RAA$ at $\pt$ of 2--3~$\gev/c$.
\item[VII] Cao, Qin, Bass~\cite{bass}. This model is also based on the Langevin approach. 
In addition to quasi-elastic scatterings, radiative energy loss is incorporated by treating gluon radiation as an additional force term. 
The space-time evolution of the medium is modelled using a viscous hydrodynamic simulation. 
The hadronization of heavy quarks has a contribution based on the recombination mechanism.
With respect to~\cite{bass}, the curves shown in the figure were obtained with a more recent parametrization for the nuclear shadowing of the parton distribution functions.
 This model provides a good description of the $\RAA$ data in central collisions, but it yields a value of $v_2$ significantly smaller than the 
 measured one (similarly to the WHDG and POWLANG models)
  and also underestimates the difference between the in-plane and out-of-plane suppression. 
\end{itemize}

Overall, the anisotropy is qualitatively described by the models that include both charm quark energy loss
in a geometrically anisotropic medium and mechanisms that transfer to charm quarks the elliptic flow induced 
during the system expansion. These mechanisms include collisional processes (MC@sHQ+EPOS, Coll+Rad(LPM)~\cite{gossiauxEPOS}, BAMPS~\cite{bamps}) and 
resonance scattering with hadronization via recombination (TAMU elastic~\cite{rapp2014}, UrQMD~\cite{lang}) in a hydrodynamically expanding 
QGP. Models that do not include a collective expansion of the medium
or lack a contribution to the hadronization of charm quarks from recombination with 
light quarks from the medium predict in general 
a smaller anisotropy than observed in the data.
The comparison for $\RAA$ and $v_2$ shows that it is challenging
to simultaneously describe the large suppression of D mesons in central collisions 
and their anisotropy in non-central collisions. In general, the models that are best in describing $\RAA$ 
tend to underestimate $v_2$ and the models that describe $v_2$ tend to underestimate the measured $\RAA$ at high $\pt$.
It is also worth noting that most of the calculations do reproduce the RHIC measurements
of heavy-flavour decay electron $\RAA$ and $v_2$.

\section{Summary}
\label{sec:summary}

We have presented a comprehensive set of results on the azimuthal anisotropy of charm production 
at central rapidity in Pb--Pb collisions at $\sqrtsNN=2.76~\tev$, obtained by reconstructing the decays
$\rm D^0\to K^-\pi^+$, $\rm D^+\to K^-\pi^+\pi^+$ and $\rm D^{*+}\to D^0\pi^+$.

The azimuthal anisotropy parameter $v_2$ was measured with the 
event plane, scalar product and two-particle cumulant methods, as a function 
of transverse momentum for semi-central collisions in the 30--50\% quantile of the hadronic cross section.
The measured anisotropy was found to be consistent among D meson species, as well as for the three methods.
The average $v_2$ of the three mesons in the interval $2<\pt<6~\gev/c$ is larger
than zero with a significance of $5.7\,\sigma$, combining statistical and systematic uncertainties.
With a smaller significance, a positive $v_2$ is also observed for $\pt>6~\gev/c$, likely to originate 
from a path-length dependence of the partonic energy loss.
The azimuthal anisotropy of $\Dzero$ mesons, which have larger statistical significance than $\Dplus$
and $\Dstar$, was also measured in the centrality classes 0--10\% and 10--30\%. 
For all three centrality classes, the $\Dzero$ meson $v_2$ is comparable in magnitude to that of inclusive charged particles.
An indication for
a decrease of $v_2$ towards more central collisions was observed for $3<\pt<6~\gev/c$.

The anisotropy was also quantified in terms of the $\Dzero$ meson nuclear modification factor $\RAA$, measured
in the direction of the event plane and orthogonal to it.  
For $\pt>3~\gev/c$, a stronger suppression relative to proton--proton collisions is observed in the out-of-plane direction, where the 
average path length of heavy quarks
through the medium is larger. 

The results indicate that, during the collective expansion of the medium, the interactions between its constituents and charm quarks transfer to the latter
information on the azimuthal anisotropy of the system.

The new results for $v_2$ and $\RAA$ measured in and out of the event plane, as well as previously published $\RAA$ in the most central 
collisions~\cite{aliceDRAA}, were compared with model calculations.
The anisotropy is best described by the models that include
mechanisms, like collisional energy loss, that transfer to charm quarks the elliptic flow induced 
during the system expansion.
In some of these models the charmed meson $v_2$ is further enhanced by 
charm quark recombination with 
light quarks from the medium.
However, it is challenging
for models to describe simultaneously the large suppression of D mesons in central collisions 
and their anisotropy in non-central collisions. 
The results reported in this article provide important constraints on the mechanisms of heavy-quark energy loss and 
on the transport properties of the expanding medium produced in
high-energy heavy-ion collisions.

%%%
%==========================================================%
%==================ACKNOWLEDGEMENTS========================%
%==========================================================%

\medskip

%%%%%%%% acknowledgements
\newenvironment{acknowledgement}{\relax}{\relax}
\begin{acknowledgement}
\section*{Acknowledgements}

The ALICE Collaboration would like to thank all its engineers and technicians for their invaluable contributions to the construction of the experiment and the CERN accelerator teams for the outstanding performance of the LHC complex.
The ALICE Collaboration gratefully acknowledges the resources and support provided by all Grid centres and the Worldwide LHC Computing Grid (WLCG) collaboration.
The ALICE Collaboration would like to thank the authors of the theoretical calculations for providing their results.
The ALICE Collaboration acknowledges the following funding agencies for their support in building and
running the ALICE detector:
State Committee of Science,  World Federation of Scientists (WFS)
and Swiss Fonds Kidagan, Armenia,
Conselho Nacional de Desenvolvimento Cient\'{\i}fico e Tecnol\'{o}gico (CNPq), Financiadora de Estudos e Projetos (FINEP),
Funda\c{c}\~{a}o de Amparo \`{a} Pesquisa do Estado de S\~{a}o Paulo (FAPESP);
National Natural Science Foundation of China (NSFC), the Chinese Ministry of Education (CMOE)
and the Ministry of Science and Technology of China (MSTC);
Ministry of Education and Youth of the Czech Republic;
Danish Natural Science Research Council, the Carlsberg Foundation and the Danish National Research Foundation;
The European Research Council under the European Community's Seventh Framework Programme;
Helsinki Institute of Physics and the Academy of Finland;
French CNRS-IN2P3, the `Region Pays de Loire', `Region Alsace', `Region Auvergne' and CEA, France;
German BMBF and the Helmholtz Association;
General Secretariat for Research and Technology, Ministry of
Development, Greece;
Hungarian OTKA and National Office for Research and Technology (NKTH);
Department of Atomic Energy and Department of Science and Technology of the Government of India;
Istituto Nazionale di Fisica Nucleare (INFN) and Centro Fermi -
Museo Storico della Fisica e Centro Studi e Ricerche "Enrico
Fermi", Italy;
MEXT Grant-in-Aid for Specially Promoted Research, Ja\-pan;
Joint Institute for Nuclear Research, Dubna;
%Korea Foundation for International Cooperation of Science and Technology (KICOS);
National Research Foundation of Korea (NRF);
CONACYT, DGAPA, M\'{e}xico, ALFA-EC and the EPLANET Program
(European Particle Physics Latin American Network);
Stichting voor Fundamenteel Onderzoek der Materie (FOM) and the Nederlandse Organisatie voor Wetenschappelijk Onderzoek (NWO), Netherlands;
Research Council of Norway (NFR);
Polish Ministry of Science and Higher Education;
National Science Centre, Poland;
 Ministry of National Education/Institute for Atomic Physics and CNCS-UEFISCDI - Romania;
Ministry of Education and Science of Russian Federation, Russian
Academy of Sciences, Russian Federal Agency of Atomic Energy,
Russian Federal Agency for Science and Innovations and The Russian
Foundation for Basic Research;
Ministry of Education of Slovakia;
Department of Science and Technology, South Africa;
CIEMAT, EELA, Ministerio de Econom\'{i}a y Competitividad (MINECO) of Spain, Xunta de Galicia (Conseller\'{\i}a de Educaci\'{o}n),
CEA\-DEN, Cubaenerg\'{\i}a, Cuba, and IAEA (International Atomic Energy Agency);
Swedish Research Council (VR) and Knut $\&$ Alice Wallenberg
Foundation (KAW);
Ukraine Ministry of Education and Science;
United Kingdom Science and Technology Facilities Council (STFC);
The United States Department of Energy, the United States National
Science Foundation, the State of Texas, and the State of Ohio.

\end{acknowledgement}

%==========================================================%
%==================BIBLIOGRAPHY============================%
%==========================================================%

%
\newpage
%
%\input{}               %%%%%%%%%%% put your appendices here
%
%%%%%%%%% appendix with author list
\appendix
\section{The ALICE Collaboration}
\label{app:collab}

% Collaboration: CERN-LHC-ALICE
% Generation Date is 2014/Apr/30

% How to use:
%%%%%%%%% appendix with author list
%\appendix
%\section{The ALICE Collaboration}
%\label{app:collab}
%\input{authors-list.tex}  %%%%%%% get the latest version before submitting

\begingroup
\small
\begin{flushleft}
B.~Abelev\Irefn{org69}\And
J.~Adam\Irefn{org37}\And
D.~Adamov\'{a}\Irefn{org77}\And
M.M.~Aggarwal\Irefn{org81}\And
M.~Agnello\Irefn{org105}\textsuperscript{,}\Irefn{org88}\And
A.~Agostinelli\Irefn{org26}\And
N.~Agrawal\Irefn{org44}\And
Z.~Ahammed\Irefn{org124}\And
N.~Ahmad\Irefn{org18}\And
I.~Ahmed\Irefn{org15}
% ...
\renewcommand\labelenumi{\textsuperscript{\theenumi}~}

\section*{Affiliation notes}
\renewcommand\theenumi{\roman{enumi}}
\begin{Authlist}
\item \Adef{0}Deceased
\item \Adef{idp1101856}{Also at: St. Petersburg State Polytechnical University}
\item \Adef{idp3004384}{Also at: Department of Applied Physics, Aligarh Muslim University, Aligarh, India}
\item \Adef{idp3687184}{Also at: M.V. Lomonosov Moscow State University, D.V. Skobeltsyn Institute of Nuclear Physics, Moscow, Russia}
\item \Adef{idp3930784}{Also at: University of Belgrade, Faculty of Physics and "Vin\v{c}a" Institute of Nuclear Sciences, Belgrade, Serbia}
\item \Adef{idp4236384}{Permanent Address: Permanent Address: Konkuk University, Seoul, Korea}
\item \Adef{idp4782160}{Also at: Institute of Theoretical Physics, University of Wroclaw, Wroclaw, Poland}
\item \Adef{idp5707472}{Also at: University of Kansas, Lawrence, KS, United States}
\end{Authlist}

\section*{Collaboration Institutes}
\renewcommand\theenumi{\arabic{enumi}~}
\begin{Authlist}

\item \Idef{org1}A.I. Alikhanyan National Science Laboratory (Yerevan Physics Institute) Foundation, Yerevan, Armenia
\item \Idef{org2}Benem\'{e}rita Universidad Aut\'{o}noma de Puebla, Puebla, Mexico
\item \Idef{org3}Bogolyubov Institute for Theoretical Physics, Kiev, Ukraine
\item \Idef{org4}Bose Institute, Department of Physics and Centre for Astroparticle Physics and Space Science (CAPSS), Kolkata, India
\item \Idef{org5}Budker Institute for Nuclear Physics, Novosibirsk, Russia
\item \Idef{org6}California Polytechnic State University, San Luis Obispo, CA, United States
\item \Idef{org7}Central China Normal University, Wuhan, China
\item \Idef{org8}Centre de Calcul de l'IN2P3, Villeurbanne, France
\item \Idef{org9}Centro de Aplicaciones Tecnol\'{o}gicas y Desarrollo Nuclear (CEADEN), Havana, Cuba
\item \Idef{org10}Centro de Investigaciones Energ\'{e}ticas Medioambientales y Tecnol\'{o}gicas (CIEMAT), Madrid, Spain
\item \Idef{org11}Centro de Investigaci\'{o}n y de Estudios Avanzados (CINVESTAV), Mexico City and M\'{e}rida, Mexico
\item \Idef{org12}Centro Fermi - Museo Storico della Fisica e Centro Studi e Ricerche ``Enrico Fermi'', Rome, Italy
\item \Idef{org13}Chicago State University, Chicago, USA
\item \Idef{org14}Commissariat \`{a} l'Energie Atomique, IRFU, Saclay, France
\item \Idef{org15}COMSATS Institute of Information Technology (CIIT), Islamabad, Pakistan
\item \Idef{org16}Departamento de F\'{\i}sica de Part\'{\i}culas and IGFAE, Universidad de Santiago de Compostela, Santiago de Compostela, Spain
\item \Idef{org17}Department of Physics and Technology, University of Bergen, Bergen, Norway
\item \Idef{org18}Department of Physics, Aligarh Muslim University, Aligarh, India
\item \Idef{org19}Department of Physics, Ohio State University, Columbus, OH, United States
\item \Idef{org20}Department of Physics, Sejong University, Seoul, South Korea
\item \Idef{org21}Department of Physics, University of Oslo, Oslo, Norway
\item \Idef{org22}Dipartimento di Fisica dell'Universit\`{a} 'La Sapienza' and Sezione INFN Rome, Italy
\item \Idef{org23}Dipartimento di Fisica dell'Universit\`{a} and Sezione INFN, Cagliari, Italy
\item \Idef{org24}Dipartimento di Fisica dell'Universit\`{a} and Sezione INFN, Trieste, Italy
\item \Idef{org25}Dipartimento di Fisica dell'Universit\`{a} and Sezione INFN, Turin, Italy
\item \Idef{org26}Dipartimento di Fisica e Astronomia dell'Universit\`{a} and Sezione INFN, Bologna, Italy
\item \Idef{org27}Dipartimento di Fisica e Astronomia dell'Universit\`{a} and Sezione INFN, Catania, Italy
\item \Idef{org28}Dipartimento di Fisica e Astronomia dell'Universit\`{a} and Sezione INFN, Padova, Italy
\item \Idef{org29}Dipartimento di Fisica `E.R.~Caianiello' dell'Universit\`{a} and Gruppo Collegato INFN, Salerno, Italy
\item \Idef{org30}Dipartimento di Scienze e Innovazione Tecnologica dell'Universit\`{a} del  Piemonte Orientale and Gruppo Collegato INFN, Alessandria, Italy
\item \Idef{org31}Dipartimento Interateneo di Fisica `M.~Merlin' and Sezione INFN, Bari, Italy
\item \Idef{org32}Division of Experimental High Energy Physics, University of Lund, Lund, Sweden
\item \Idef{org33}Eberhard Karls Universit\"{a}t T\"{u}bingen, T\"{u}bingen, Germany
\item \Idef{org34}European Organization for Nuclear Research (CERN), Geneva, Switzerland
\item \Idef{org35}Faculty of Engineering, Bergen University College, Bergen, Norway
\item \Idef{org36}Faculty of Mathematics, Physics and Informatics, Comenius University, Bratislava, Slovakia
\item \Idef{org37}Faculty of Nuclear Sciences and Physical Engineering, Czech Technical University in Prague, Prague, Czech Republic
\item \Idef{org38}Faculty of Science, P.J.~\v{S}af\'{a}rik University, Ko\v{s}ice, Slovakia
\item \Idef{org39}Frankfurt Institute for Advanced Studies, Johann Wolfgang Goethe-Universit\"{a}t Frankfurt, Frankfurt, Germany
\item \Idef{org40}Gangneung-Wonju National University, Gangneung, South Korea
\item \Idef{org41}Gauhati University, Department of Physics, Guwahati, India
\item \Idef{org42}Helsinki Institute of Physics (HIP), Helsinki, Finland
\item \Idef{org43}Hiroshima University, Hiroshima, Japan
\item \Idef{org44}Indian Institute of Technology Bombay (IIT), Mumbai, India
\item \Idef{org45}Indian Institute of Technology Indore, Indore (IITI), India
\item \Idef{org46}Institut de Physique Nucl\'eaire d'Orsay (IPNO), Universit\'e Paris-Sud, CNRS-IN2P3, Orsay, France
\item \Idef{org47}Institut f\"{u}r Informatik, Johann Wolfgang Goethe-Universit\"{a}t Frankfurt, Frankfurt, Germany
\item \Idef{org48}Institut f\"{u}r Kernphysik, Johann Wolfgang Goethe-Universit\"{a}t Frankfurt, Frankfurt, Germany
\item \Idef{org49}Institut f\"{u}r Kernphysik, Westf\"{a}lische Wilhelms-Universit\"{a}t M\"{u}nster, M\"{u}nster, Germany
\item \Idef{org50}Institut Pluridisciplinaire Hubert Curien (IPHC), Universit\'{e} de Strasbourg, CNRS-IN2P3, Strasbourg, France
\item \Idef{org51}Institute for Nuclear Research, Academy of Sciences, Moscow, Russia
\item \Idef{org52}Institute for Subatomic Physics of Utrecht University, Utrecht, Netherlands
\item \Idef{org53}Institute for Theoretical and Experimental Physics, Moscow, Russia
\item \Idef{org54}Institute of Experimental Physics, Slovak Academy of Sciences, Ko\v{s}ice, Slovakia
\item \Idef{org55}Institute of Physics, Academy of Sciences of the Czech Republic, Prague, Czech Republic
\item \Idef{org56}Institute of Physics, Bhubaneswar, India
\item \Idef{org57}Institute of Space Science (ISS), Bucharest, Romania
\item \Idef{org58}Instituto de Ciencias Nucleares, Universidad Nacional Aut\'{o}noma de M\'{e}xico, Mexico City, Mexico
\item \Idef{org59}Instituto de F\'{\i}sica, Universidad Nacional Aut\'{o}noma de M\'{e}xico, Mexico City, Mexico
\item \Idef{org60}iThemba LABS, National Research Foundation, Somerset West, South Africa
\item \Idef{org61}Joint Institute for Nuclear Research (JINR), Dubna, Russia
\item \Idef{org62}Korea Institute of Science and Technology Information, Daejeon, South Korea
\item \Idef{org63}KTO Karatay University, Konya, Turkey
\item \Idef{org64}Laboratoire de Physique Corpusculaire (LPC), Clermont Universit\'{e}, Universit\'{e} Blaise Pascal, CNRS--IN2P3, Clermont-Ferrand, France
\item \Idef{org65}Laboratoire de Physique Subatomique et de Cosmologie, Universit\'{e} Grenoble-Alpes, CNRS-IN2P3, Grenoble, France
\item \Idef{org66}Laboratori Nazionali di Frascati, INFN, Frascati, Italy
\item \Idef{org67}Laboratori Nazionali di Legnaro, INFN, Legnaro, Italy
\item \Idef{org68}Lawrence Berkeley National Laboratory, Berkeley, CA, United States
\item \Idef{org69}Lawrence Livermore National Laboratory, Livermore, CA, United States
\item \Idef{org70}Moscow Engineering Physics Institute, Moscow, Russia
\item \Idef{org71}National Centre for Nuclear Studies, Warsaw, Poland
\item \Idef{org72}National Institute for Physics and Nuclear Engineering, Bucharest, Romania
\item \Idef{org73}National Institute of Science Education and Research, Bhubaneswar, India
\item \Idef{org74}Niels Bohr Institute, University of Copenhagen, Copenhagen, Denmark
\item \Idef{org75}Nikhef, National Institute for Subatomic Physics, Amsterdam, Netherlands
\item \Idef{org76}Nuclear Physics Group, STFC Daresbury Laboratory, Daresbury, United Kingdom
\item \Idef{org77}Nuclear Physics Institute, Academy of Sciences of the Czech Republic, \v{R}e\v{z} u Prahy, Czech Republic
\item \Idef{org78}Oak Ridge National Laboratory, Oak Ridge, TN, United States
\item \Idef{org79}Petersburg Nuclear Physics Institute, Gatchina, Russia
\item \Idef{org80}Physics Department, Creighton University, Omaha, NE, United States
\item \Idef{org81}Physics Department, Panjab University, Chandigarh, India
\item \Idef{org82}Physics Department, University of Athens, Athens, Greece
\item \Idef{org83}Physics Department, University of Cape Town, Cape Town, South Africa
\item \Idef{org84}Physics Department, University of Jammu, Jammu, India
\item \Idef{org85}Physics Department, University of Rajasthan, Jaipur, India
\item \Idef{org86}Physik Department, Technische Universit\"{a}t M\"{u}nchen, Munich, Germany
\item \Idef{org87}Physikalisches Institut, Ruprecht-Karls-Universit\"{a}t Heidelberg, Heidelberg, Germany
\item \Idef{org88}Politecnico di Torino, Turin, Italy
\item \Idef{org89}Purdue University, West Lafayette, IN, United States
\item \Idef{org90}Pusan National University, Pusan, South Korea
\item \Idef{org91}Research Division and ExtreMe Matter Institute EMMI, GSI Helmholtzzentrum f\"ur Schwerionenforschung, Darmstadt, Germany
\item \Idef{org92}Rudjer Bo\v{s}kovi\'{c} Institute, Zagreb, Croatia
\item \Idef{org93}Russian Federal Nuclear Center (VNIIEF), Sarov, Russia
\item \Idef{org94}Russian Research Centre Kurchatov Institute, Moscow, Russia
\item \Idef{org95}Saha Institute of Nuclear Physics, Kolkata, India
\item \Idef{org96}School of Physics and Astronomy, University of Birmingham, Birmingham, United Kingdom
\item \Idef{org97}Secci\'{o}n F\'{\i}sica, Departamento de Ciencias, Pontificia Universidad Cat\'{o}lica del Per\'{u}, Lima, Peru
\item \Idef{org98}Sezione INFN, Bari, Italy
\item \Idef{org99}Sezione INFN, Bologna, Italy
\item \Idef{org100}Sezione INFN, Cagliari, Italy
\item \Idef{org101}Sezione INFN, Catania, Italy
\item \Idef{org102}Sezione INFN, Padova, Italy
\item \Idef{org103}Sezione INFN, Rome, Italy
\item \Idef{org104}Sezione INFN, Trieste, Italy
\item \Idef{org105}Sezione INFN, Turin, Italy
\item \Idef{org106}SSC IHEP of NRC Kurchatov institute, Protvino, Russia
\item \Idef{org107}SUBATECH, Ecole des Mines de Nantes, Universit\'{e} de Nantes, CNRS-IN2P3, Nantes, France
\item \Idef{org108}Suranaree University of Technology, Nakhon Ratchasima, Thailand
\item \Idef{org109}Technical University of Split FESB, Split, Croatia
\item \Idef{org110}The Henryk Niewodniczanski Institute of Nuclear Physics, Polish Academy of Sciences, Cracow, Poland
\item \Idef{org111}The University of Texas at Austin, Physics Department, Austin, TX, USA
\item \Idef{org112}Universidad Aut\'{o}noma de Sinaloa, Culiac\'{a}n, Mexico
\item \Idef{org113}Universidade de S\~{a}o Paulo (USP), S\~{a}o Paulo, Brazil
\item \Idef{org114}Universidade Estadual de Campinas (UNICAMP), Campinas, Brazil
\item \Idef{org115}University of Houston, Houston, TX, United States
\item \Idef{org116}University of Jyv\"{a}skyl\"{a}, Jyv\"{a}skyl\"{a}, Finland
\item \Idef{org117}University of Liverpool, Liverpool, United Kingdom
\item \Idef{org118}University of Tennessee, Knoxville, TN, United States
\item \Idef{org119}University of Tokyo, Tokyo, Japan
\item \Idef{org120}University of Tsukuba, Tsukuba, Japan
\item \Idef{org121}University of Zagreb, Zagreb, Croatia
\item \Idef{org122}Universit\'{e} de Lyon, Universit\'{e} Lyon 1, CNRS/IN2P3, IPN-Lyon, Villeurbanne, France
\item \Idef{org123}V.~Fock Institute for Physics, St. Petersburg State University, St. Petersburg, Russia
\item \Idef{org124}Variable Energy Cyclotron Centre, Kolkata, India
\item \Idef{org125}Vestfold University College, Tonsberg, Norway
\item \Idef{org126}Warsaw University of Technology, Warsaw, Poland
\item \Idef{org127}Wayne State University, Detroit, MI, United States
\item \Idef{org128}Wigner Research Centre for Physics, Hungarian Academy of Sciences, Budapest, Hungary
\item \Idef{org129}Yale University, New Haven, CT, United States
\item \Idef{org130}Yonsei University, Seoul, South Korea
\item \Idef{org131}Zentrum f\"{u}r Technologietransfer und Telekommunikation (ZTT), Fachhochschule Worms, Worms, Germany
\end{Authlist}
\endgroup

  %%%%%%% get the latest version before submitting

\end{document}